\documentclass[superscriptaddress,twocolumn,reprint]{revtex4-1}

\usepackage{array}		
\usepackage{amsmath}    
\usepackage{supertabular}
\usepackage{graphicx}   
\usepackage{verbatim}   
\usepackage{color}      
\usepackage{subfigure}  
\usepackage{epspdfconversion} 
\usepackage{hyperref}   

\begin{document}

\title{ Nonequilibrium Thermodynamics of  Porous Electrodes }  
\author{Todd R. Ferguson}
\affiliation{Department of Chemical Engineering, Massachusetts Institute of Technology}
\author{Martin Z. Bazant}
\affiliation{Department of Chemical Engineering, Massachusetts Institute of Technology}
\affiliation{Department of Mathematics, Massachusetts Institute of Technology}
\date{\today}  

\begin{abstract}
We reformulate and extend porous electrode theory for non-ideal active materials, including those capable of phase transformations.  Using principles of non-equilibrium thermodynamics, we relate the cell voltage, ionic fluxes, and Faradaic charge-transfer kinetics to the variational electrochemical potentials of ions and electrons.   The Butler-Volmer exchange current is consistently expressed in terms of the activities of the reduced, oxidized and transition states, and the activation overpotential is defined relative to the local Nernst potential.  We also apply mathematical bounds on effective diffusivity to estimate porosity and tortuosity corrections. The theory is illustrated for a Li-ion battery with active solid particles described by a Cahn-Hilliard phase-field model.  Depending on the applied current and porous electrode properties, the dynamics can be limited by electrolyte transport, solid diffusion and phase separation, or intercalation kinetics. In phase-separating porous electrodes, the model predicts narrow reaction fronts, mosaic instabilities and voltage fluctuations at low current, consistent with recent experiments, which could not be described by existing porous electrode models. 
\end{abstract}

\maketitle

\section{Introduction}

Modeling is a key component of any design process.  An accurate model allows one to interpret experimental data, identify rate limiting steps and predict system behavior, while providing a deeper understanding of the underlying physical processes. In systems engineering,  empirical models with fitted parameters are often used for design and control, but it is preferable, whenever possible, to employ models based on microscopic physical or geometrical parameters, which can be more easily interpreted and optimized.  

In the case of electrochemical energy storage devices, such as batteries, fuel cells, and supercapacitors, the systems approach is illustrated by equivalent circuit models, which are widely used in conjunction with impedance spectroscopy to fit and predict cell performance and degradation. This approach is limited, however, by the difficulty in unambiguously interpreting fitted circuit elements and in making predictions for the nonlinear response to large operating currents.  There is growing interest, therefore, in developing physics-based porous electrode models and applying them for battery optimization and control~\cite{ramadesigan2012}. Quantum mechanical computational methods have demonstrated the possibility of predicting bulk material properties, such as open circuit potential and solid diffusivity, from first principles~\cite{ceder1998}, but coarse-grained continuum models are needed to describe the many length and time scales of interfacial reactions and multiphase, multicomponent transport phenomena. 

Mathematical models could play a crucial role in guiding the development of new intercalation materials, electrode microstructures, and battery architectures, in order to meet the competing demands in power density and energy density for different envisioned applications, such as electric vehicles or renewable (e.g. solar, wind) energy storage. 
Porous electrode theory, pioneered by J. Newman and collaborators, provides the standard modeling framework for battery simulations today~\cite{newman_book}.  As reviewed in the next section, this approach has been developed for over half a century and applied successfully to many battery systems. The treatment of the active material, however, remains rather simple, and numerous parameters are often needed to fit experimental data. 

In porous electrode theory for Li-ion batteries, transport is modeled via volume averaged conservation equations \cite{devidts1997}.  The solid active particles are modeled as spheres, where intercalated lithium undergoes  isotropic linear diffusion ~\cite{doyle1993,doyle1996}.  For phase separating materials, such as Li$_x$FePO$_4$ (LFP),  each particle is assumed to have a spherical phase boundary that moves as a ``shrinking core", as one phase displaces the other~\cite{srinivasan2004,dargaville2010,thorat2011}. In these models, the local Nernst equilibrium potential is fitted to the global open circuit voltage of the cell, but this neglects non-uniform composition, which  makes the voltage plateau an emergent property of the porous electrode~\cite{dreyer2010,dreyer2011,bai2011,cogswell2012}. For thermodynamic consistency,  all of these phenomena should derive from common thermodynamic principles and cannot be independently fitted to experimental data. The open circuit voltage reflects the activity of intercalated ions, which in turn affects ion transport in the solid phase and Faradaic reactions involving ions in the electrolyte phase~\cite{10.626,acr2012}.  

In this paper, we extend porous electrode theory to non-ideal active materials, including those capable of phase transformations.  Our starting point is a general phase-field theory of ion intercalation kinetics developed by our group over the past five years~\cite{acr2012,singh2008,burch2008wave,large_acis,burch2009,bai2011}, which has recently led to a quantitative understanding of phase separation dynamics in LFP nanoparticles~\cite{cogswell2012}.  The ionic fluxes in all phases are related to electrochemical potential gradients~\cite{lai2011a,large_acis}, consistent with non-equilibrium thermodynamics~\cite{degroot_book,kom}.
For thermodynamic consistency, the  Faradaic reaction rate is also related to electrochemical potential differences between the oxidized, reduced, and transition states, leading to a generalized Butler-Volmer equation~\cite{acr2012} suitable for phase-separating materials.  These elements are integrated in a general porous electrode theory, where the active material is described by a Cahn-Hilliard phase-field model~\cite{kom,nauman2001}, as in nanoscale simulations of Li-ion battery materials~\cite{han2004,garcia2005,singh2008,burch2008wave,burch2009,tang2011,kang2009,bai2011,cogswell2012}. This allows us to describe the non-equilibrium thermodynamics of porous battery electrodes in terms of  well established physical principles for ion intercalation in nanoparticles.

\section{ Background }

\subsection{ Mathematical Modeling of Porous Electrodes } 

We begin by briefly reviewing volume-averaged porous electrode theory, which has been the standard approach in battery modeling for the past 50 years, in order to highlight similarities and differences with our approach. The earliest attempts to formulate porous electrode models ~\cite{ksenzhek1956,euler1960} related current density distributions to macroscopic properties such as porosity, average surface area per volume, and effective conductivity, and capacitive charging was added in transmission line models~\cite{delevie1963}. Sixty years ago, the seminal work  Newman and Tobias~\cite{newman1962} first described the effects of concentration variations on kinetics and introduced the well-known mass conservation equations for porous electrodes, which form the basis for modern battery modeling.  Extensive literature surveys are available by Newman and coauthors~\cite{newman1975,newman_book} for work up to the 1990s.

Here, we only draw attention to some specific papers and recent developments that set the stage for our theoretical approach. Perhaps the earliest use of concepts from non-equilibrium thermodynamics in porous electrode theory was by Ksenzhek, who incorporated concentrated solution theory in the transport equations inside a porous electrode, and referred to gradients in electrochemical potential as the driving force for transport \cite{ksenzhek1964}. This is the fundamental postulate of linear irreversible thermodynamics in chemical physics~\cite{degroot_book} and materials science~\cite{kom}, and it has also recently been applied to electrochemical systems~\cite{jamnik2001,lai2005,lai2010,lai2011a,lai2011b,biesheuvel2007,kilic2007a,kilic2007b,olesen2010,bazant2011} and electrokinetic phenomena~\cite{large_acis,large_new,biesheuvel2005,storey2008}.  Although concentrated solution theory is widely applied to batteries~\cite{newman_book}, the thermodynamic driving force for transport has only recently been connected to the battery voltage~\cite{lai2010,lai2011a,lai2011b} and Faradaic reaction kinetics~\cite{acr2012,bai2011,cogswell2012}.

Porous electrode theories make a number of underlying assumptions regarding properties of the cell that can be critical to performance. For example, an early paper of Grens \cite{grens1970} showed that the assumption of constant conductivity for the electron conducting phase is usually valid, while the assumption of constant electrolyte concentration, often used for mathematical convenience, is only valid over a narrow range of operating conditions. These concepts are extended here to volume averaging over solid reaction products undergoing phase transformations (which further narrows the range of validity of porous electrode models to exclude mosaic instabilities among discrete particles in a representative continuum volume element).

Our work also focuses on the nonlinear dynamics of porous electrodes, which could only be addressed as computer power improved. Early work focused on steady state \cite{newman1962,gurevich1967}, mostly at small (linearized Butler-Volmer) or large (Tafel regime) overpotentials~\cite{grens1965}, or transient response for small sinusoidal perturbations (impedance)~\cite{rangarajan1968} or fast kinetics~\cite{pollard1980}. Similar to our motivation below,  Atlung \emph{et al.}\cite{atlung1979} investigated the dynamics of solid solution (i.e. intercalation) electrodes for different time scales with respect to the limiting current, although without considering configurational entropy and chemical potentials as in this work. 

As computers and numerical methods advanced, so did simulations of porous electrodes, taking into account various nonlinearities in transport and reaction kinetics.  West \emph{et al.} first demonstrated the use of numerical methods to simulate discharge of a porous TiS$_2$ electrode (without the separator) in the typical case of electrolyte transport limitation \cite{west1982}.  Doyle, Fuller and Newman first simulated Li-ion batteries under constant current discharge with full Butler-Volmer kinetics for two porous electrodes and a porous separator  \cite{doyle1993,fuller1994,doyle1996}.  
These papers are of great importance in the field, as they developed the first complete simulations of lithium-ion batteries and solidified the role of porous electrode theory in modeling these systems.  The same theoretical framework has been applied to many other types of cells, such as lithium-sulfur~\cite{kumaresan2008} and LFP~\cite{srinivasan2004,dargaville2010} batteries, with particular success for lithium polymer batteries at high discharge rates 

Battery models invariably assume electroneutrality, but diffuse charge in porous electrodes has received increasing attention over the past decade, driven by applications in energy storage and desalination.  The effects of double-layer capacitance in a  porous electrode were originally considered using only linearized low-voltage models \cite{johnson1971,tiedemann1975}, which are equivalent to transmission line circuits~\cite{delevie1963,eikerling2005}. Recently, the full nonlinear dynamics of capacitive charging and salt depletion have been analyzed and simulated in both flat  \cite{bazant2004,olesen2010} and porous \cite{biesheuvel2010} electrodes. The combined effects of electrostatic capacitance and pseudo-capacitance due to Faradaic reactions have also been incorporated in porous electrode theory~\cite{biesheuvel2011,biesheuvel2012}, using Frumkin-Butler-Volmer kinetics~\cite{biesheuvel2009}. These models have been successfully used to predict the nonlinear dynamics of capacitive desalination by porous carbon electrodes~\cite{biesheuvel2011b,porada2012}.  Although we do not consider double layers in our examples below (as is typical for battery discharge), it would be straightforward to integrate these recent models into our theoretical framework based on non-equilibrium thermodynamics~\cite{bazant2011,acr2012}. 

Computational and experimental advances have also been made to study porous electrodes at the microstructural level and thus test the formal volume-averaging, which underlies macroscopic continuum models.  Garcia et al. performed finite-element simulations of ion transport in typical porous microstructures for Li-ion batteries~\cite{garcia2005}, and  Garcia and Chang simulated hypothetical inter-penetrating  3D battery architectures at the particle level~\cite{garcia2007}. Recently, Smith, Garcia and Horn analyzed the effects of microstructure on battery performance for various sizes and shapes of particles in a Li$_{1-x}$C$_6$/Li$_x$CoO$_2$ cell~\cite{smith2009}.  The study used 3D image reconstruction of a real battery microstructure by focused ion beam milling, which has led to detailed studies of  microstructural effects  in porous electrodes \cite{thorat2009, thiedmann2011, kehrwald2011}. In this paper, we will discuss mathematical bounds on effective diffusivities in porous media, which could be compared to results for actual battery microstructures. Recently, it has also become possible to observe lithium ion transport at the scale in individual particles in porous Li-ion battery electrodes~\cite{balke2010,weichert2012}, which could be invaluable in testing the dynamical predictions of new mathematical models.

\subsection{ Lithium Iron Phosphate }

The discovery of LFP as a cathode material by the Goodenough group in 1997 has had a large and unexpected impact on the battery field, which provides the motivation for our work.  LFP was first thought to be a low-power material, and it demonstrated poor capacity at room temperature. \cite{padhi1997}  The capacity has since been improved via conductive coatings and the formation of nanoparticles. \cite{ravet2001,huang2001}, and the rate capability has been improved in similar ways \cite{hsu2004,malik2010}. With high carbon loading to circumvent electronic conductivity limitations, LFP nanoparticles can now be discharged in 10 seconds~\cite{kang2009}. Off-stoichiometric phosphate glass coatings contribute to this high rate, not only in LFP, but also in LiCoO$_2$ ~\cite{sun2011}. 

It has been known since its discovery that LFP is a phase separating material, as evidenced by a flat voltage plateau in the open circuit voltage~\cite{padhi1997,tarascon2001}. There are a wide variety of battery materials with multiple stable phases at different states of charge~\cite{huggins_book}, but Li$_x$FePO$_4$ has a particularly strong tendency for phase separation, with a miscibility gap (voltage plateau) spanning across most of the range from $x=0$ to $x=1$ at room temperature. Padhi et al. first depicted phase separation inside LFP particles schematically as a  ``shrinking core'' of one phase being replaced by an outer shell of the other phase during charge/discharge cycles~\cite{padhi1997}.  Srinivasan and Newman encoded this concept in a porous electrode theory of the  LFP cathode with spherical active particles, containing spherical shrinking cores. \cite{srinivasan2004}  Recently, Dargaville and Farrell have expanded this approach to predict active material utilization in LFP electrodes. \cite{dargaville2010}  Thorat \emph{et al.} have also used the model to gain insight into rate-limiting mechanisms inside LFP cathodes. \cite{thorat2011}  

To date, the shrinking-core porous electrode model is the only model to successfully fit the galvanostatic discharge of an LFP electrode, but the results are not fully satisfactory. Besides neglecting the microscopic physics of phase separation, the model relies on fitting a concentration-dependent solid diffusivity, whose inferred values are orders of magnitude smaller than {\it ab initio} simulations \cite{morgan2004,malik2010} or impedance measurements~\cite{pasquali2008}. More consistent values of the solid diffusivity have since been obtained by different models attempting to account for anisotropic phase separation with elastic coherency strain. \cite{zhu2010}  Most troubling for the shrinking core picture, however, is the direct observation of phase boundaries with very different orientations. In 2006, Chen, Song, and Richardson published images showing the orientation of the phase interface aligned with iron phosphate planes and reaching the active facet of the particle. \cite{chen2006}  This observation was supported by experiments of  Delmas \emph{et al.}, who suggested a ``domino-cascade model" for the intercalation process inside LFP \cite{delmas2008}. With further experimental evidence for anisotropic phase morphologies~\cite{oyama2012,weichert2012}, it has become clear that a new approach is needed to capture the non-equilibrium thermodynamics of this material.

\subsection{ Phase-Field Models }

Phase-field models are widely used to describe phase transformations and microstructural evolution in materials science~\cite{kom,chen2002}, but they are relatively new to electrochemistry.  In 2004, Guyer, Boettinger, Warren and McFadden~\cite{guyer2004a,guyer2004b} first modeled the sharp electrode/electrolyte interface with a continuous phase field varying between stable values $0$ and $1$, representing the liquid electrolyte and solid metal phases. As in phase-field models of dendritic solidification~\cite{karma1994,boettinger1996,boettinger2000,boettinger2002}, they used a simple quartic function to model a double-welled homogeneous free energy. They described the kinetics of electrodeposition~\cite{guyer2004b} (converting ions in the electrolyte to solid metal) by Allen-Cahn-type kinetics~\cite{allen1979,chen2002}, linear in the thermodynamic driving force, but did not make connections with the Butler-Volmer equation. Several groups have used this approach to model dendritic electrodeposition   and related processes~\cite{assadi2006,shibuta2006,pong2007}. Also in 2004, Han, Van der Ven and Ceder~\cite{han2004}  first applied the Cahn-Hilliard equation\cite{cahn1958,cahn1959-1,cahn1959-2,cahn1961,kom,chen2002} to the diffusion of intercalated lithium ions in LFP, albeit without modeling reaction kinetics.

Building on these advances, Bazant developed a general theory of charge-transfer and Faradaic reaction kinetics in concentrated solutions and solids based on non-equilibrium thermodynamics~\cite{acr2012,10.95,10.626}, suitable for use with phase-field models. The exponential Tafel dependence of the current  on the overpotential, defined in terms of the variational chemical potentials, was first reported in 2007 by Singh, Ceder and Bazant~\cite{singh2008,singh2007}, but with spurious pre-factor, corrected by Burch~\cite{burch_thesis,burch2009}. The model was used to predict ``intercalation waves" in small, reaction-limited LFP nanoparticles in 1D~\cite{singh2008}, 2D~\cite{burch2008wave}, and 3D~\cite{tang2011}, thus providing a mathematical description of the domino cascade phenomenon~\cite{delmas2008}.  The complete electrochemical phase-field theory, combining the Cahn-Hilliard with Butler-Volmer kinetics and the cell voltage, appeared in 2009 lectures notes~\cite{10.95,10.626} and was applied to LFP nanoparticles~\cite{bai2011,cogswell2012}.  

The new theory has led to a quantitative understanding of intercalation dynamics in single nanoparticles of LFP. 
Bai, Cogswell and Bazant~\cite{bai2011} generalized the Butler-Volmer equation using variational chemical potentials (as derived in the supporting information) and used it to develop a mathematical theory of the suppression of phase separation in LFP nanoparticles with increasing current. This phenomenon, which helps to explain the remarkable performance of nano-LFP, was also suggested by Malik and Ceder based on bulk free energy calculations~\cite{malik2011}, but the theory shows that it is entirely controlled by Faradaic reactions at the particle surface~\cite{bai2011,cogswell2012}. Cogswell and Bazant~\cite{cogswell2012} have shown that including elastic coherency strain in the model leads to a quantitative theory of phase morphology and lithium solubility. Experimental data for different particles sizes and temperatures can be fitted with only two parameters (the gradient penalty and regular solution parameter, defined below). 

The goal of the present work is to combine the phase-field theory of ion intercalation in nanoparticles  with classical porous electrode theory to arrive at a general mathematical framework for non-equilibrium thermodynamics of porous electrodes. Our work was first presented at the Fall Meeting of the Materials Research Society in  2010 and again at the Electrochemical Society Meetings in Montreal and Boston in 2011. Around the same time, Lai and Ciucci were thinking along similar lines~\cite{lai2010,lai2011b} and published an important reformulation of Newman's porous electrode theory based non-equilibrium thermodynamics~\cite{lai2011a}, but they did not make any connections with phase-field models or phase transformations at the macroscopic electrode scale. Their treatment of reactions also differs from Bazant's theory of generalized Butler-Volmer or Marcus kinetics~\cite{acr2012,10.626,10.95}, with a thermodynamically consistent description of the transition state in charge transfer.

In this paper, we develop a variational thermodynamic description of electrolyte transport, electron transport, electrochemical kinetics, and phase separation, and we apply to Li-ion batteries  in what appears to be the first mathematical theory and computer simulations of {\it macroscopic phase transformations in porous electrodes}. Simulations of discharge into a cathode consisting of multiple phase-separating particles interacting via an electrolyte reservoir at constant chemical potential were reported by Burch~\cite{burch_thesis}, who observed ``mosaic instabilities", where particles transform one-by-one at low current. This phenomenon was elegantly described by Dreyer et al. in terms of a (theoretical and experimental) balloon model, which helps to explain the noisy voltage plateau and zero-current voltage gap in slow charge/discharge cycles of porous LFP electrodes~\cite{dreyer2010,dreyer2011}.  These studies, however, did not account for electrolyte transport and associated macroscopic gradients in porous electrodes undergoing phase transformations, which are the subject of this work. To do this, we must reformulate Faradaic reaction kinetics for concentrated solutions, consistent with the Cahn-Hilliard equation for ion intercalation and Newman's porous electrode theory for the electrolyte.

\section{General Theory of Reactions and Transport in Concentrated Solutions}

In this section, we begin with a general theory of reaction rates based on non-equilibrium thermodynamics and transition state theory. We then expand the model to treat transport in concentrated solutions (i.e. solids).  Finally, we show that this concentrated solution model collapses to Fickian diffusion in the dilute limit.  For more details and examples, see Refs. ~\cite{acr2012,10.626}.

\subsection{General Theory of Reaction Rates}

The theory begins with the diffusional chemical potential of species $i$,
\begin{equation}
\mu_i = k_B T \ln c_i  +  \mu_i^{ex} = k_BT \ln a_i 
\end{equation}
where $c_i$ is the concentration, $a_i$ is the absolute chemical activity, $\mu_i^{ex}  = k_BT \ln \gamma_i$ is the excess chemical potential in a concentrated solution, and $\gamma_i$ is the activity coefficient ($a_i = \gamma_i c_i$). In linear irreversible thermodynamics (LIT)~\cite{degroot_book,prigogine_book,kom}, the flux of species $i$ is proportional to its chemical potential gradient, as discussed below.

\begin{figure}[htp]
\centering	
	\includegraphics[width=3in,keepaspectratio]{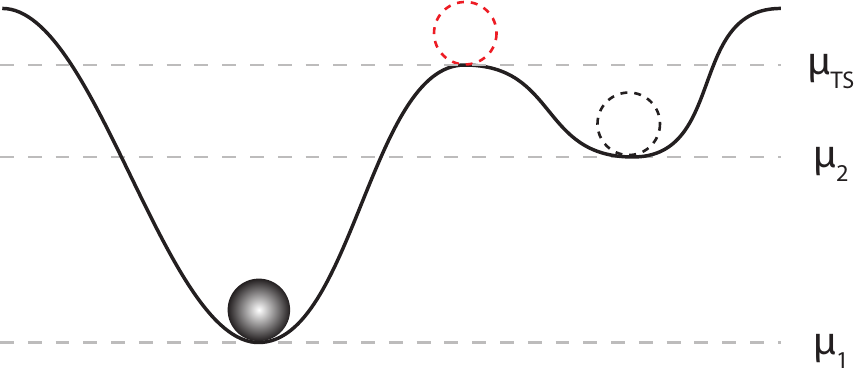}
\caption{\textbf{Typical reaction energy landscape}.  The set of atoms involved in the reaction travels through a transition state as it passes from one state to the other in a landscape of total excess chemical potential as a function of the atomic coordinates. }
\label{reactionratefig01}
\end{figure}

In a thermodynamically consistent formulation of reaction kinetics~\cite{10.626,sekimoto_book}, therefore, the reaction complex explores  a landscape of {\it excess} chemical potential $\mu^{ex}(x)$ between local minima $\mu_1^{ex}$ and $\mu_2^{ex}$ with transitions over an activation barrier $\mu^{ex}_{\ddag}$, as shown in Figure \ref{reactionratefig01}. For long-lived states with rare transitions ($\mu^{ex}_\ddag - \mu_i^{ex}\gg k_BT$), the net reaction rate is given by 
\begin{eqnarray}
R &=& R_{1\to 2} - R_{2\to 1} \nonumber \\
&=& \nu \left[ e^{-(\mu^{ex}_\ddag-\mu_1)/k_B T} - e^{-(\mu^{ex}_\ddag-\mu_2)/k_B T} \right] 
\label{netrxn1}\\
&=& \frac{\nu(a_1-a_2)}{\gamma_\ddag}  \nonumber
\end{eqnarray}
which automatically satisfies the De Donder relation~\cite{sekimoto_book}, 
\begin{equation}
\mu_1 - \mu_2 = k_B T \ln\left( \frac{R_{1\to 2}}{R_{2\to 1}}\right).
\end{equation}
The frequency prefactor $\nu$ depends on generalized force constants at the saddle point and in one minimum (e.g. state 1, with a suitable shift of $\mu^{ex}_\ddag$) as in  Kramers' escape formula~\cite{kramers1940,vankampen_book} and classical transition state theory~\cite{vineyard1957,kaxiras_book}.

For the general reaction, 
\begin{equation}
S_1 = \sum_i s_i M_i \to  \sum_j s_j M_j = S_2,
\end{equation} 
the activities, $a_1 = \prod_i a_i^{s_i}$ and $a_2 = \prod_j a_j^{s_j}$, are equal in equilibrium, and the forward and backward reactions are in detailed balance ($R=0$). The equilibrium constant is thus the ratio of the reactant to product activity coefficients:
\begin{eqnarray}
K&=& \frac{c_2}{c_1}=\frac{\prod_j c_j^{s_j}}{\prod_i c_i^{s_i}} 
=  \frac{\prod_i \gamma_i^{s_i}}{\prod_j \gamma_j^{s_j}} \nonumber \\
&=& \frac{\gamma_1}{\gamma_2} = e^{(\mu^{ex}_1-\mu^{ex}_2)/k_BT} = e^{- \Delta G^{ex}/k_BT}
\end{eqnarray}
where $\Delta G^{ex}$ is the {\it excess} free energy change per reaction.
In order to describe reaction kinetics, however, we also need a model for the transition state activity coefficient $\gamma_\ddag$, in (\ref{netrxn1}).

The subtle difference between total and excess chemical potential is often overlooked in chemical kinetics.  Lai and Ciucci~\cite{lai2010,lai2011a,lai2011b}, who also recently applied non-equilibrium thermodynamics to batteries,  postulate a Faradaic reaction rate based on a barrier of total (not excess) chemical potential. The equilibrium condition (Nernst equation) is the same, but the rate (exchange current) is  different and does not consistently treat the transition state. We illustrate this point by deriving solid diffusion and Butler-Volmer kinetics from the same reaction formalism.

\subsection{General Theory of Transport in Solids and Concentrated Solutions}

\begin{figure}[htp]
\centering	
	\includegraphics[width=3in,keepaspectratio]{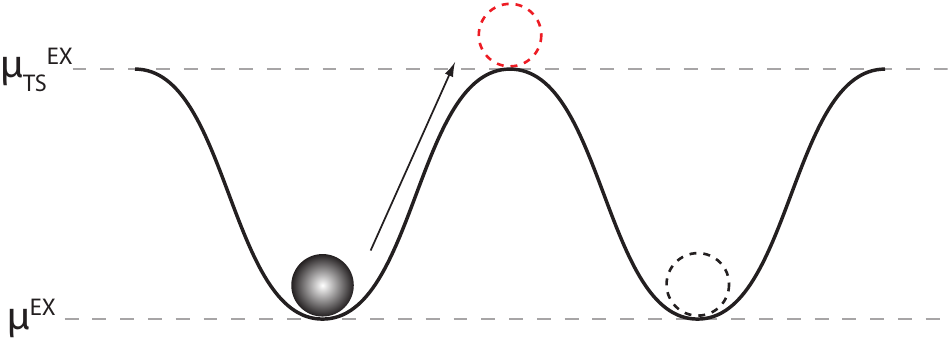}
\caption{\textbf{Typical diffusion energy landscape}. The same principles for reactions can also be applied to solid diffusion, where the diffusing molecule  explores a landscape of excess chemical potential, hopping by thermal activation between nearly equivalent local minima.  }
\label{diffusion02}
\end{figure}

In solids, atoms (or  more generally, molecules) fluctuate in long-lived states near local free energy minima and occasionally move through a transition state to a neighboring well of similar free energy.  In a crystal, the wells correspond to lattice sites, but similar concepts also apply to amorphous solids.   Figure \ref{diffusion02} demonstrates this picture of diffusion and shows an energy (or excess chemical potential) landscape for an atom moving through a medium. Tracer diffusion of individual atoms consists of thermally activated jumps over some distance between sites with an average  ``first passage time"~\cite{vankampen_book} between these transitions, $\tau$, which is the inverse of the mean transition rate per reaction event above. Using the general thermodynamic theory of reaction rates above for the activated diffusion process, the  time between transitions scales as   
\begin{equation}
\tau = \tau_o \exp\left(\frac{\mu_{\ddag}^{ex}-\mu^{ex}}{k_BT}\right).
\label{avgtime1}
\end{equation}
The tracer diffusivity, $D$, is then the mean square distance divided by the mean transition  time,
\begin{equation}
D = \frac{\left(\Delta x\right)^2}{2\tau} = \frac{\left(\Delta x\right)^2}{2\tau_o}\left(\frac{\gamma}{\gamma_{\ddag}}\right)=D_o\left(\frac{\gamma}{\gamma_{\ddag}}\right),
\label{Ddef1}
\end{equation}
where $D_o$ is the tracer diffusivity in the dilute-solution limit.

\subsubsection{Diffusivity of an Ideal Solid Solution}

\begin{figure}[htp]
\centering	
	\includegraphics[width=3in,keepaspectratio]{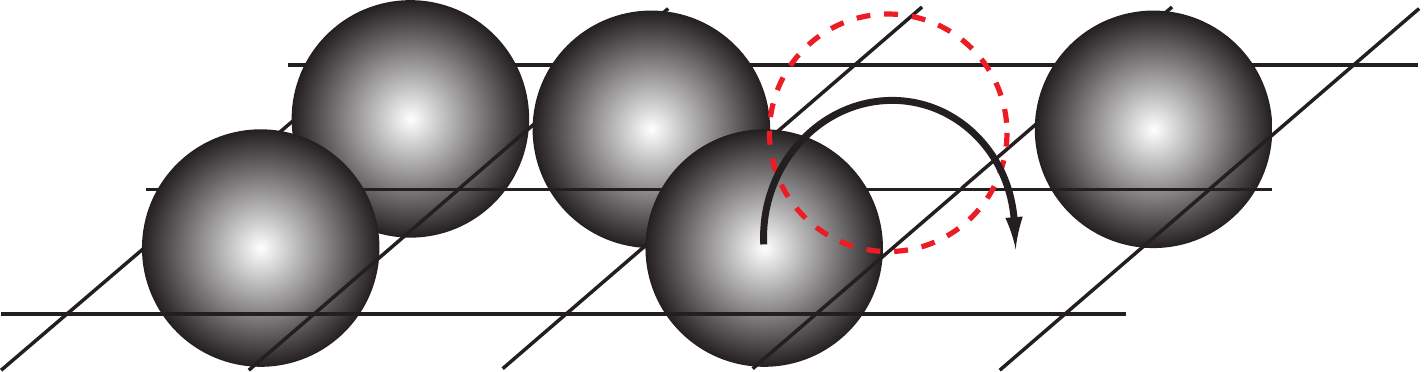}
\caption{\textbf{Lattice gas model for diffusion}.  The atoms are assigned a constant excluded volume by occupying sites on a grid.  Atoms can only jump to an open space, and the transition state (red dashed circle) requires two empty spaces.}
\label{diffusion03lg}
\end{figure}

To model an ideal solid solution, we consider a lattice gas model for the configurational entropy, which accounts for finite volume effects in the medium, and neglect any direct atom-atom interactions which contribute to the enthalpy.  Figure \ref{diffusion03lg} illustrates this model.
The chemical potential for an atom in an ideal solid solution is
\begin{equation}
\mu = k_BT\ln\left(\frac{\tilde{c}}{1-\tilde{c}}\right)+\mu^o,
\label{chempotss}
\end{equation}
where \(\mu^o\) is the chemical potential of the reference state and $\tilde{c} = c/c_{max}$ is the dimensionless concentration.  
The excluded volume of an atom is one lattice site.  However, the transition state requires two available sites, effectively doubling the excluded volume contribution to the chemical potential.  Using the definition of the activity coefficient, $\mu = k_BT\ln a = k_BT\ln\left(c\gamma\right)$, we obtain the activity coefficients of the atom in the site, and in the activated state,
\begin{eqnarray}
\gamma &=& \left(1-\frac{c}{c_{max}}\right)^{-1}\exp\left(-\frac{\mu_{min}}{k_BT}\right),
\label{activcoeffmin} \\
\gamma_{\ddag} &=& \left(1-\frac{c}{c_{max}}\right)^{-2}\exp\left(-\frac{\mu_{\ddag}}{k_BT}\right). \label{activcoeffts} 
\end{eqnarray}
Inserting these two activity coefficients into Equation \ref{Ddef1}, the diffusivity, $D$, is 
\begin{equation}
D = D_o\left(1-\frac{c}{c_{max}}\right).
\label{idssD}
\end{equation}
This diffusivity is for an ideal solid solution with a finite number of lattice sites available for atoms ~\cite{nauman2001}.  As the lattice sites fill, the diffusivity of an atom goes to zero, since the atom is unable to move as it is blocked by other atoms on the lattice.  

\subsubsection{Concentrated Solution Theory Derivation}

\begin{figure}[htp]
\centering	
	\includegraphics[width=3in,keepaspectratio]{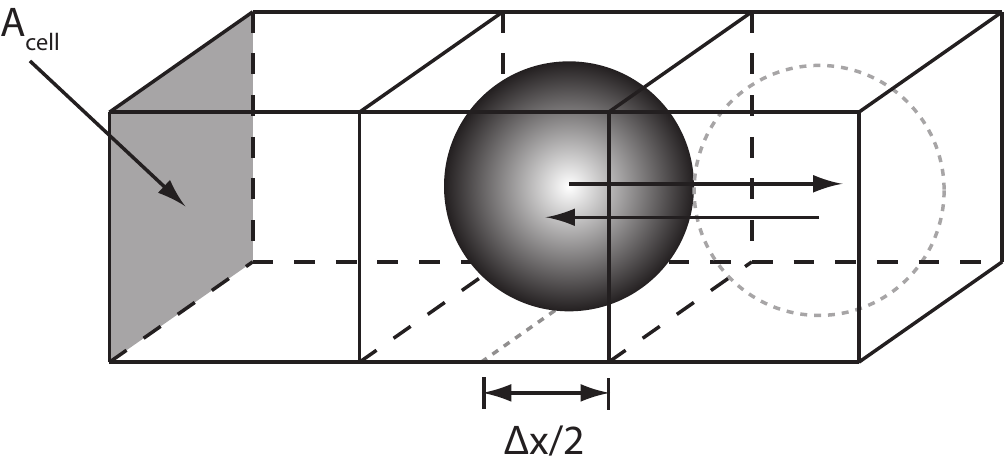}
\caption{\textbf{Diffusion through a solid}.  The flux is given by the reaction rate across the area of the cell, $A_{cell}$. In this lattice model, atoms move between available sites.}
\label{diffusion04}
\end{figure}

Here we will derive the general form of concentrated solution theory, which postulates that the flux can be modeled as
\begin{equation}
\textbf{F} = -Mc\nabla\mu,
\label{cstflux1}
\end{equation}
where $M$ is the mobility.  Let us consider the scenario in Figure \ref{diffusion03lg}, where an atom is sitting in an energy well.  This atom's energy fluctuates on the order of $k_BT$ until it has enough energy to overcome some energy barrier that exists between the two states.  Figure \ref{diffusion04} demonstrates this in one dimension.  The flux, $\textbf{F}$, is 
\begin{equation}
\textbf{F}_i = \frac{R}{A}\textbf{e}_i,
\label{fluxdef1}
\end{equation}
where $\textbf{e}_i$ is a coordinate vector in the $i$ direction and $\textbf{F}_i$ is the flux in the $i$ direction.  

We see that the atom's chemical potential is a function of location, as concentrations and therefore chemical potentials, will vary with position.  Let's define the right side of the page as the positive x-direction.  Using our previously defined form of the reaction rate in Equation (\ref{netrxn1}), we can substitute this into Equation (\ref{fluxdef1}).  However, we need an expression for the barrier-less reaction rate.  This comes from the barrier-less diffusion time in Equation (\ref{avgtime1}).  The barrier-less reaction rate should be equivalent to the inverse of two times the barrier-less diffusion time, 
\begin{equation}
R_o = \frac{1}{2\tau_o}.
\label{Roderv1}
\end{equation}
The one half comes from the probability the atom travels in the positive x direction.  Plugging this into Equation (\ref{fluxdef1}) along with Equation (\ref{netrxn1}), and considering the fact that our chemical potential is a function of position, we obtain
\begin{eqnarray}
\textbf{F}_x & =&  \frac{1}{2\tau_oA_{cell}\gamma_{\ddag}}\left[\exp\left(\tilde{\mu}(x)-\frac{\Delta x}{2}\frac{\partial\tilde{\mu}(x)}{\partial x}\right)   \right. \nonumber \\
& & - \left. \exp\left(\tilde{\mu}(x)+\frac{\Delta x}{2}\frac{\partial\tilde{\mu}(x)}{\partial x}\right)\right],
\label{fluxsimp1}
\end{eqnarray}
where $\tilde{\mu}(x)$ denotes the chemical potential scaled by the thermal voltage, $k_BT$.  Next, we assume that the atom is close to equilibrium.  That is, the difference in chemical potential between the states is small.  This allows us to linearize Equation (\ref{fluxsimp1}).  Linearizing the equation yields
\begin{equation}
\textbf{F}_x = -\frac{a(x)}{\tau_oA_{cell}\gamma_{\ddag}}\left(\frac{\Delta x}{2}\right)\frac{\partial\tilde{\mu}(x)}{\partial x},
\label{fluxsimp2}
\end{equation}
where $a(x)$ is the activity as a function of position.  This can be simplified to $a(x) = V\gamma(x)c(x)$.  Plugging this into Equation (\ref{fluxsimp2}), using our definition of the diffusivity, $D$, from Equation (\ref{Ddef1}), and the Einstein relation, which states that $M=D/k_BT$, we obtain the flux as predicted by concentrated solution theory in the x-dimension.  We can easily expand this to other dimensions.  Doing so, we obtain the form of the flux proposed by concentrated solution theory,
\begin{equation}
\textbf{F} = -Mc\nabla\mu,
\label{fluxsimp3}
\end{equation}
where $c = c(x,y,z)$.  Taking the dilute limit, as $c\rightarrow 0$, and using the definition of chemical potential, $\mu = k_BT\ln a$, where $a = \gamma c$ and $\gamma=1$ (dilute limit), we obtain Fick's Law from Equation (\ref{fluxsimp3}),
\begin{equation}
\textbf{F}=-D\nabla c.
\label{fickslaw}
\end{equation}

\section{Characterization of Porous Media}

In batteries, the electrodes are typically composites consisting of active material (e.g. graphite in the anode, iron phosphate in the cathode), conducting material (e.g. carbon black), and binder.  The electrolyte penetrates the pores of this solid matrix.  This porous electrode is advantageous because it substantially increases the available active area of the electrode.  However, this type of system, which can have variations in porosity (i.e. volume of electrolyte per volume of the electrode) and loading percent of active material throughout the volume, presents difficulty in modeling.  To account for the variation in electrode properties, various volume averaging methods for the electrical conductivity and transport properties in the electrode are employed.  In this section, we will give a brief overview of modeling the conductivity and transport of a heterogeneous material, consisting of two or more materials with different properties~\cite{milton2002,torquato2002,stauffer_book,sahimi2003}

\subsection{Electrical Conductivity of the Porous Media}

\begin{figure}[htp]
\centering	
	\includegraphics[width=3in,keepaspectratio]{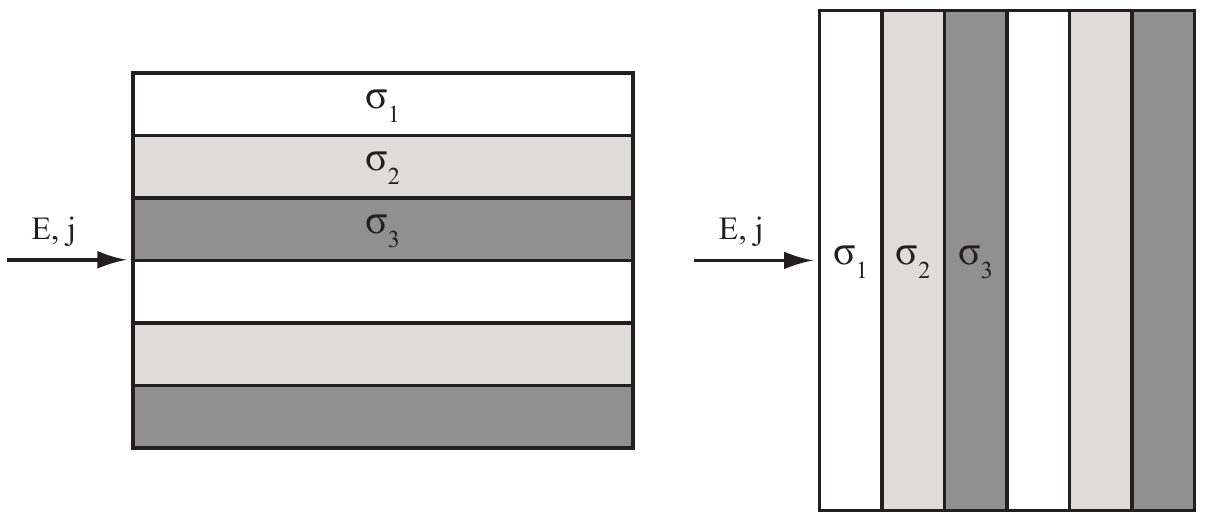}
\caption{  \textbf{Wiener bounds on  the effective conductivity of a two-phase anisotropic material}.  The left figure demonstrates the upper conductivity limit achieved by stripes aligned with the field, which act like resistors in parallel.  The right figure demonstrates the lower bound with the materials arranged in transverse stripes to act like resistors in series. }
\label{upperloweraniso1}
\end{figure}

To characterize the electrical conductivity of the porous media, we will consider rigorous mathematical bounds over all possible microstructures with the same volume fractions of each component.  First we consider a general anisotropic material as shown in Figure \ref{upperloweraniso1}, in which case the conductivity bounds, due to Wiener, are attained by simple microstructures with parallel stripes of the different materials~\cite{torquato2002}. 
The left image in Figure \ref{upperloweraniso1} represents the different materials as resistors in parallel, which produces the lowest possible resistance and the upper limit of the conductivity of the heterogeneous material.  The right image represents the materials as resistors in series, which produces the highest possible resistance, or lower limit of the conductivity.  
These limits are referred to as the upper and lower Wiener bounds, respectively.  Let $\Phi_i$ be the volume fraction of material $i$.  For the upper Wiener bound, attained by stripes parallel to the current, the  effective conductivity is simply the arithmetic mean of the individual conductivities, weighted by their volume fractions,   
\begin{equation}
\overline{\sigma}_{max} = \langle\sigma\rangle = \sum_i\Phi_i\sigma_i.
\label{upperwiener1}
\end{equation}
The lower Wiener bound is attained by stripes perpendicular to the current, and the effective conductivity is a weighted harmonic mean of the individual conductivities, as for resistors in parallel,
\begin{equation}
\overline{\sigma}_{min} = \langle\sigma^{-1}\rangle^{-1} = \frac{1}{\sum_i\frac{\Phi_i}{\sigma_i}}.
\label{lowerwiener1}
\end{equation}
For a general anisotropic material, the effective conductivity, $\overline{\sigma}$, must lie within the Wiener bounds,
\begin{equation}
\langle\sigma^{-1}\rangle^{-1}\leq \overline{\sigma} \leq \langle\sigma\rangle.
\label{wienerbounds1}
\end{equation}

\begin{figure}[htp]
\centering	
	\includegraphics[width=3in,keepaspectratio]{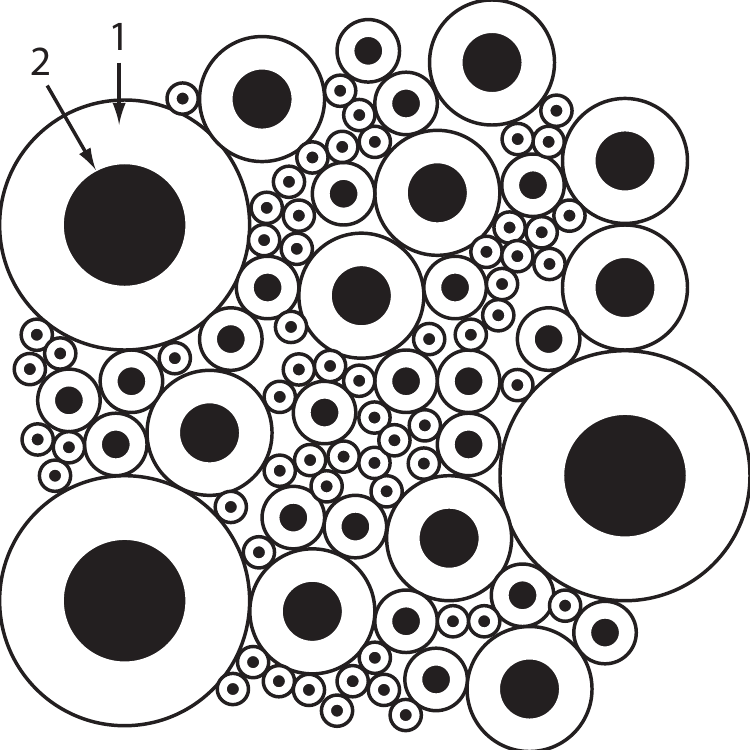}
\caption{\textbf{ Hashin-Shtrikman bounds on the effective conductivity of a two-phase isotropic material}.  Isotropic random composite of space-filling coated spheres which attain the bounds.  The white represents material with conductivity $\sigma_1$ and the black represents material with conductivity $\sigma_2$.  Maximum conductivity is achieved when $\sigma_1>\sigma_2$ and minimum conductivity is obtained when $\sigma_2>\sigma_1$. 
The volume fractions $\Phi_1$ and $\Phi_2$ are the same. }
\label{hsmodel1}
\end{figure}

There are tighter bounds on the possible effective conductivity of isotropic media, which have no preferred direction, due to Hashin and Shtrikman (HS)~\cite{torquato2002}.  There are a number of microstructures which attain the HS bounds, such as a space-filling set of concentric circles or spheres, whose radii are chosen to set the given volume fractions of each material. The case of two components is shown in Figure \ref{hsmodel1}.   
The HS lower bound on conductivity is attained by ordering the individual materials so as to place the highest conductivity at the core and the lowest conductivity in the outer shell, of each particle. For the HS  upper bound, the ordering is reversed, and the lowest conductivity material is buried in the core of each particle, while the highest conductivity is in the outer shell, forming a percolating network across the system.

\begin{figure}[htp]
\centering	
	\includegraphics[width=3in,keepaspectratio]{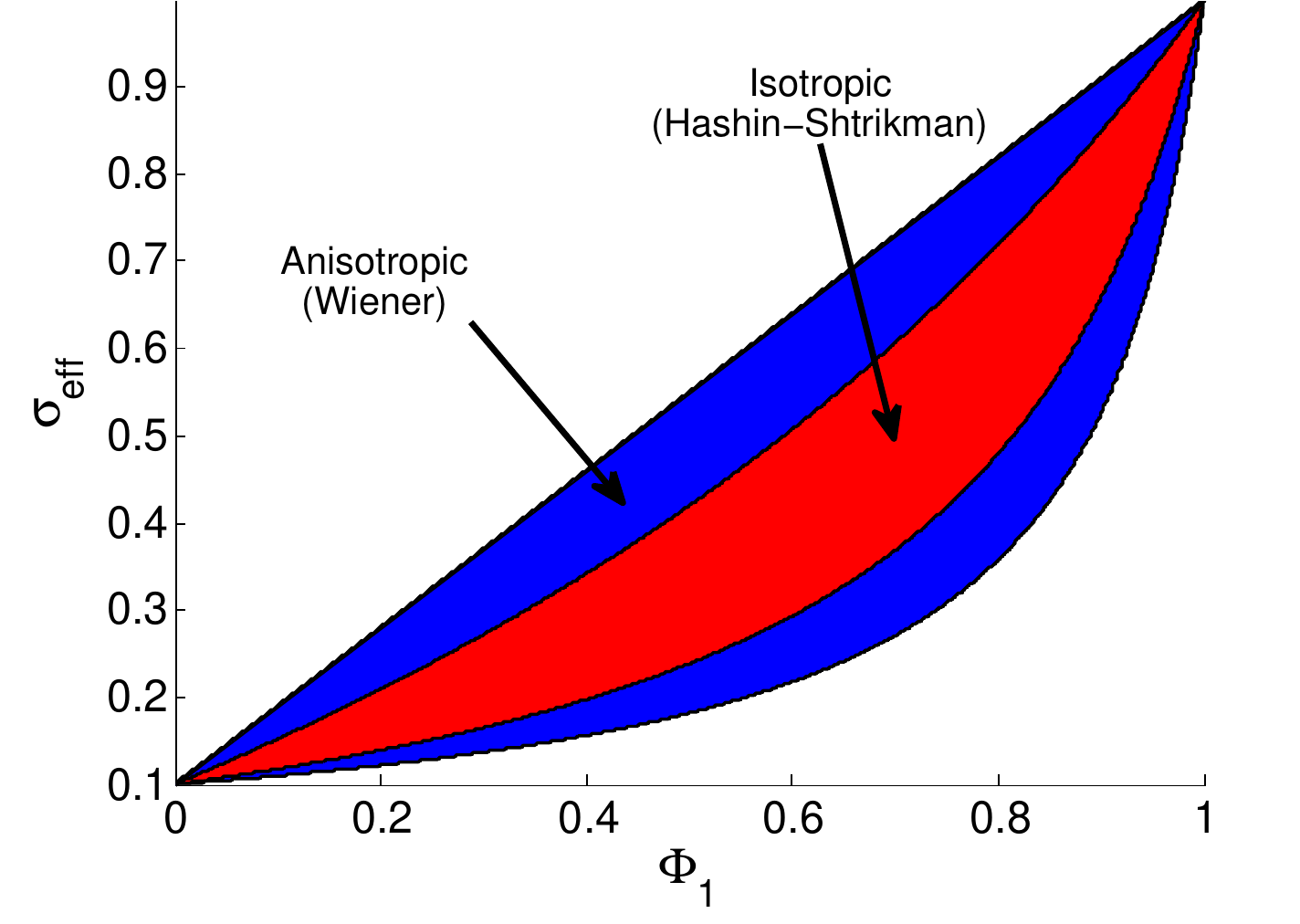}
\caption{\textbf{Conductivity bounds for two-phase composites versus volume fraction}.  The above figure shows the Wiener bounds (blue) for an anisotropic two component material and Hashin-Shtrikman bounds (red) for an isotropic two component material versus the volume fraction of material 1.  The conductivities used to produce the figure are $\sigma_1=1$ and $\sigma_2=0.1$.}
\label{condbounds1}
\end{figure}

For the case of two components, where $\sigma_1>\sigma_2$, the HS conductivity bounds for an isotropic two-component material in $d$ dimensions are 
\begin{equation}
\langle\sigma\rangle-\frac{\left(\sigma_1-\sigma_2\right)^2\Phi_1\Phi_2}
{\langle\tilde{\sigma}\rangle+\sigma_2\left(d-1\right)}\leq\overline{\sigma}\leq
\langle\sigma\rangle-\frac{\left(\sigma_1-\sigma_2\right)^2\Phi_1\Phi_2}
{\langle\tilde{\sigma}\rangle+\sigma_1\left(d-1\right)},
\label{hsbounds1}
\end{equation}
where
\[
\langle\sigma\rangle=\Phi_1\sigma_1+\Phi_2\sigma_2
\]
and
\[
\langle\tilde{\sigma}\rangle=\Phi_1\sigma_2+\Phi_2\sigma_1.
\]
The Wiener and Hashin-Shtrikman bounds above provide us with possible ranges for isotropic and anisotropic media with two components.  Figure \ref{condbounds1} gives the Wiener and Hashin-Shtrikman bounds for two materials, with conductivities of 1.0 and 0.1.

Next, we consider ion transport in porous media.  Ion transport in porous media often consists of a solid phase, which has little to no ionic conductivity (i.e. slow or no diffusion) permeated by an electrolyte phase which has very high ionic conductivity (i.e. fast diffusion).  In the next section, we will compare different models for effective porous media properties.

\subsection{ Conduction in Porous Media}

\begin{figure}[htp]
\centering	
	\includegraphics[width=3in,keepaspectratio]{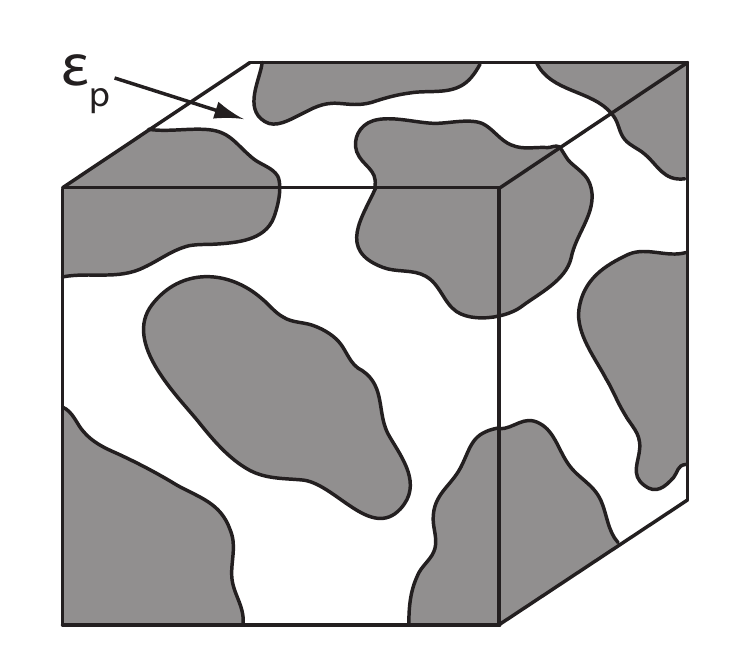}
\caption{\textbf{Example of a porous volume}.  This is an example of a typical porous volume.  A mixture of solid particles is permeated by an electrolyte.  The porosity, $\epsilon_p$, is the volume of electrolyte as a fraction of the volume of the cube.}
\label{porvolume1}
\end{figure}

For the case of ion transport in porous media, there is an electrolyte phase, which has a non-zero diffusivity, and the solid phase, through which transport is very slow (essentially zero compared to the electrolyte diffusivity).  Here, we consider the pores (electrolyte phase) and give the solid matrix a zero conductivity.  The volume fraction of phase 1 (the pores), $\Phi_1$, is the porosity:
\[
\Phi_1 = \epsilon_p\mbox{, }
\sigma_1 = \sigma_p.
\]
The conductivity for all other phases is zero.  This reduces the Wiener (anisotropic) and Hashin-Shtrikman (isotropic) lower bounds to zero.  Figure (\ref{porvolume1}) demonstrates a typical volume of a porous medium. 

\begin{figure}[htp]
\centering	
	\includegraphics[width=3in,keepaspectratio]{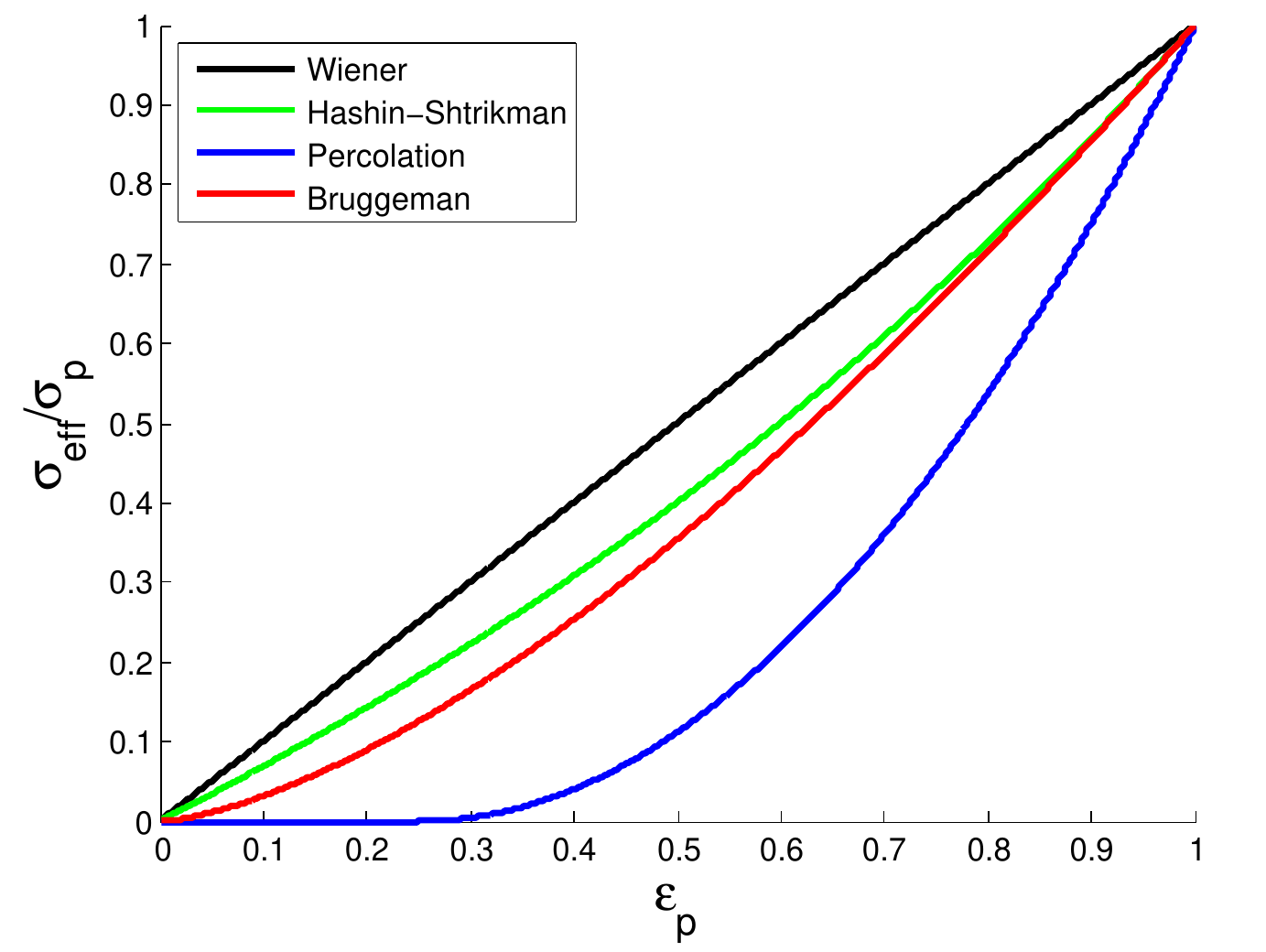}
\caption{\textbf{Various models for effective conductivity in 3D}.  This figure demonstrates the effective conductivity (scaled by the pore conductivity) using Wiener bounds, Hashin-Shtrikman bounds, a percolation model, and the Bruggeman formula. The percolation model uses a critical porosity of $\epsilon_c=0.25$.}
\label{condbounds2}
\end{figure}

In porous electrode models for batteries~\cite{doyle1993,fuller1994,srinivasan2004}, the empirical Bruggeman formula is used to relate the conductivity to the porosity, 
\begin{equation}
\overline{\sigma}_{B} = \epsilon_p^{3/2}\sigma_p
\label{bruggeman1}
\end{equation}
although it is not clear what mathematical approximation is being made. As shown in  Figure \ref{condbounds2}, the Bruggeman formula turns out to be close to (and fortunately, below) the HS upper bound, so we can see that it corresponds to a highly conducting isotropic material, similar to a core-shell microstructure with solid cores and conducting shells. This makes sense for ionic conductivity in liquid-electrolyte-soaked porous media, but not for electronic conductivity based on networks of touching particles. 

To understand the possible range of conductivity, we consider the rigorous bounds above.
If we assume the media consists of two phases ($\Phi_2=1-\epsilon_p\mbox{, }\sigma_2=0$), then the Wiener and Hashin-Shtrikman upper bounds can be simplified to 
\begin{equation}
\overline{\sigma}^{Wiener}_{max} = \Phi_1\sigma_1 = \epsilon_p\sigma_p,
\label{poroswienermax}
\end{equation}
and
\begin{equation}
\overline{\sigma}^{HS}_{max} = \sigma_p\epsilon_p\left(\frac{d-1}{d-\epsilon_p}\right). 
\label{porosHSmax}
\end{equation}
where again $d$ is the embedding dimension. The HS upper bound is attained by spherical core-shell particles with the conducting pore phase spanning the system via conducting shells on non-conducting solid cores, similar to electron-conducting coatings on active battery particles~\cite{awarkea2011}.

The lower bounds vanish because it is always possible that the conducting phase does not ``percolate", or form a continuous path, across the system.   Equivalently, the non-conducting matrix phase can percolate and block conduction. In such situations, however, the bounds are of little use, since they give no sense of the probability of finding percolating paths through a random microstructure.  For ionic conduction through the electrolyte, which permeates the matrix, percolation may not be a major issue, but for electron conduction it is essential to maintain a network of touching conducting particles (such as carbon black in a typical battery electrode)~\cite{awarkea2011}.

In statistical physics, percolation models serve to quantify the conductivity of random media due to geometrical connectivity of particles~\cite{stauffer_book,sahimi2003}. The simplest percolation models corresponds to randomly coloring a lattice of sites or bonds with a probability equal to the mean porosity and measuring the statistics of conduction through clusters of connected sites or bonds. Continuum percolation models, such as the ``swiss cheese model", correspond to randomly placing or removing overlapping particles of given shapes to form clusters. The striking general feature of such models is the existence of a critical porosity $\epsilon_c$ in the thermodynamic limit of an infinite system, below which the probability of a spanning infinite cluster is zero, and above which it is one. 
The critical point depends on the specific percolation model, and for lattice models and decreases with increasing coordination number (mean number of connected neighbors), as more paths across the system are opened.  Just above the critical point, the effective conductivity scales as a power law 
\begin{equation}
\overline{\sigma}_{perc}\sim\left(\epsilon_p-\epsilon_c\right)^{t_p}
\label{perceffectcond}
\end{equation}
where the exponent is believed to be universal for all percolation models in the same embedding dimensions and equal to $t_p=2$ in three dimensions.  A simple form to capture this behavior is   
\begin{equation}
\overline{\sigma}_{perc} \cong \begin{cases}
\sigma_p\left(\frac{\epsilon_p-\epsilon_c}{1-\epsilon_c}\right)^2 & \epsilon_c\leq\epsilon_p\leq 1 \\
0 & 0\leq\epsilon_p\leq\epsilon_c
\end{cases}.
\end{equation}

\subsection{ Diffusion in Porous Media }

We now relate the conductivity to the effective diffusivity of the porous medium.  
The porosity is the volume of the electrolyte as a fraction of the total volume.  If the porosity is assumed to be constant throughout the volume, then the area of each face of the volume is proportional to the porosity.  Also, the total mass inside the volume is given by the volume averaged concentration, $\overline{c}=\epsilon_pc$.  We begin with a mass balance on the volume,
\begin{equation}
\frac{\partial \overline{c}}{\partial t} + \nabla\cdot\mathbf{F} = 0,
\label{massbal1porvol1}
\end{equation}
where $\textbf{F}$ is the flux at the surfaces of the volume.  The net flux is
\begin{equation}
\mathbf{F} = -\overline{\sigma}_d\nabla c,
\label{netfluxeffcond1}
\end{equation}
where $c$ is the concentration in the pores and $\overline{\sigma}_d$ is the mean diffusive conductivity of the porous medium (with the same units as diffusivity, m$^2$/s), which, as the notation suggests, can be approximated or bounded by the conductivity formulae in the previous section, with $\sigma_p$ replaced by the ``free-solution" diffusivity $D_p$ within the pores.  It is important to recognize that fluxes are driven by gradients in the microscopic concentration within the pores, $c$, and not the macroscopic, volume-averaged concentration, $\bar{c}$.  Regardless of porosity fluctuations in space, at equilibrium the concentration within the pores, which determines the local chemical potential, is constant throughout the volume.

Combining Equations (\ref{massbal1porvol1}) and (\ref{netfluxeffcond1}), we get
\begin{equation}
\frac{\partial c}{\partial t} = \overline{D}\nabla^2c,
\label{massbal1porvol2}
\end{equation}
where the effective diffusivity in a porous medium, $\overline{D}$, is given by
\begin{equation}
\overline{D}=\frac{\overline{\sigma}_d}{\epsilon_p}.
\label{effectiveDiff1}
\end{equation}
The reduction of the diffusivity inside a porous medium can be interpreted as a reduction of the mean free path. The tortuosity, $\tau_p$, is often used to related the effective macroscopic diffusivity to the microscopic diffusivity within the pores,
\begin{equation}
\overline{D} = \frac{D_p}{\tau_p},
\label{Defftort1}
\end{equation}
as suggested long ago by Peterson~\cite{peterson1958}.  One must keep in mind, however, that the tortuosity is just a way of interpreting the effective diffusivity in a porous medium, which is not rigorously related to any geometrical property of the microstructure.  In Fick's Law, which involves one spatial derivative, the tortuosity can be interpreted as the ratio of an effective microscopic diffusion path length $L_p$ to the macroscopic geometrical length: $L_p = \tau_p L$, 
although it is usually not clear exactly what kind of averaging is performed over all possible paths.  Indeed, other definitions of tortuosity are also used~\cite{shen2007}.  (In particular, if the length rescaling concept is applied to the diffusion equation, which has two spatial derivatives,  then the definition $\overline{D}=D_p/\tau^2$ is more natural, but equally arbitrary.)

\begin{figure}[htp]
\centering	
	\includegraphics[width=3in,keepaspectratio]{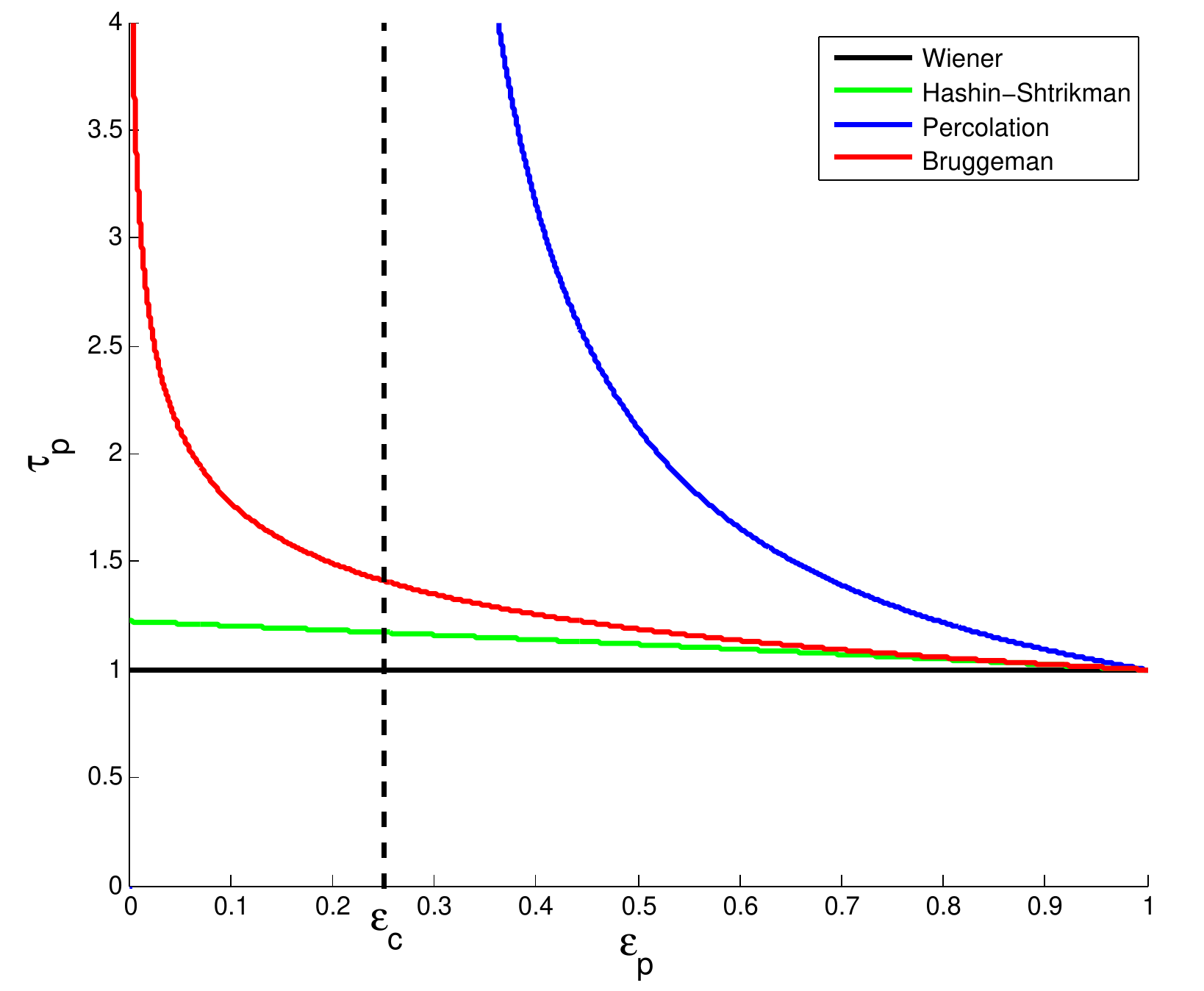}
\caption{\textbf{Tortuosity versus porosity for different effective conductivity models}.  This plot gives the tortuosity for different porosity values.  While the Wiener and Hashin-Shtrikman models produce finite tortuosities, the percolation and Bruggeman models diverge as porosity goes to zero.}
\label{tortplot1}
\end{figure}

In any case, using the definition above, the effective conductivity can be expressed as
\begin{equation}
\overline{\sigma}_d = \frac{D_p\epsilon_p}{\tau_p}
\label{effconddiff1}
\end{equation}
which allows us to interpret all the models and bounds above in terms of Peterson's tortuosity $\tau_p$.  The upper bounds on conductivity become lower bounds on tortuosity. The Wiener lower bound tortuosity for anisotropic pores is
\begin{equation}
\tau_p^{Wiener} = 1.
\label{wienertortlower1e}
\end{equation}
For the Hashin-Shtrikman model, the lower bound of the tortuosity is
\begin{equation}
\tau_p^{HS} = \frac{d-\epsilon_p}{d-1}
\label{hstortlower1}
\end{equation}
in $d$ dimensions.  The percolation model produces a piecewise function for the tortuosity, above and below the critical porosity, which is given by
\begin{equation}
\tau_p^{perc} \cong \begin{cases}
\epsilon_p\left(\frac{1-\epsilon_c}{\epsilon_p-\epsilon_c}\right)^2 & \epsilon_c\leq\epsilon_p\leq 1 \\
\infty & 0\leq\epsilon_p\leq\epsilon_c
\end{cases}
\label{perctortlower1}
\end{equation}
Note that, as the conductivity approaches zero, the tortuosity makes no physical sense as it no longer represents the extra path length.  Instead it represents the decreasing number of available percolating paths, which are the cause of the lowered conductivity.  Finally, from the Bruggeman empirical relation we get the empirical tortuosity formula,
\begin{equation}
\tau_p^{B} = \epsilon^{-1/2},
\label{wienertortlower1}
\end{equation}
which is widely used in porous electrode models for batteries, stemming from the work of J. Newman and collaborators. The different tortuosity models are plotted in Figure \ref{tortplot1}, and we note again the close comparison of the Bruggeman-Newman formula to the rigorous Hashin-Shtrikman upper bound for an isotropic porous medium.

\section{Porous Electrode Theory}

\subsection{ Conservation Equations }
Using the principles laid out in the first section of this paper on concentrated solution theory, the Porous Electrode Theory equations will be derived using mass and charge conservation combined with the Nernst-Planck Equation and a modified form of the Butler-Volmer Equation.  The derivation will present the equations and how their properties have deep ties to the thermodynamics of the system.  Then, the equations will be non-dimensionalized and scaled appropriately using characteristic time and length scales in the system.
\subsubsection{Mass and Charge Conservation}
We begin with the definition of flux based on concentrated solution theory.  Assuming the system is close to equilibrium, the mass flux is
\begin{equation}
\textbf{N}_i = -M_ic_i\nabla\mu_i,
\label{gencstflux1}
\end{equation}
where $M_i$ is the mobility of species $i$, $c_i$ is the concentration of species $i$, and $\mu_i$ is the chemical potential of species $i$.  The conservation equation for concentration is given by the divergence of the flux,
\begin{equation}
\frac{\partial c_i}{\partial t} = -\nabla\cdot\textbf{N}_i-R_i.
\label{genconseqn1}
\end{equation}
It is important to note that $R_i$ is the volumetric consumption of species $i$.  In order to express this conservation equation in a form that is relevant to electrochemical systems, we must first postulate a suitable form of the chemical potential.  We begin with the standard definition of the chemical potential including the activity contribution, then include electrostatic effects to obtain
\begin{equation}
\mu_i = k_BT\ln\left(a_i\right)+z_ie\phi.
\label{electrochempot1}
\end{equation}
This chemical potential can be inserted into Equation (\ref{gencstflux1}).  If the activity of the electrolyte is available from experimental values, then this form of the flux facilitates its use.  However, diffusivities are typically given as a function of concentration.  Simplifying Equation (\ref{gencstflux1}) using Equation (\ref{electrochempot1}) for the chemical potential yields the Nernst-Planck Equation,
\begin{equation}
\textbf{N}_{i,\pm} = -D_{chem,i}\nabla c_{i} \mp \frac{ez_{i}}{k_BT}D_{i}c_{i}\nabla\phi,
\label{nernstplanck1}
\end{equation}
where $D_{chem,i}$ is the chemical diffusivity of species $i$, which is defined as
\begin{equation}
D_{chem,i} = D_i\left(1+\frac{\partial\ln\gamma_{i}}{\partial\ln c_{i}}\right).
\label{chemdiff1}
\end{equation}
The dilute limit diffusivity, $D_i$, can also have concentration dependence.  Above, $\gamma_i$ is the activity coefficient, and $\phi$ is the potential.  The charge of the species is $z_i$, which is treated as the absolute value.

For the bulk electrolyte, the electroneutrality approximation will be used.  This approximation assumes that the double layers are thin, which is a reasonable approximation when there is no depletion in the electrolyte. (For porous electrode modeling including double layer effects, see Refs. \cite{biesheuvel2010,biesheuvel2011,biesheuvel2012}.)  The electroneutrality approximation assumes
\begin{equation}
\rho = z_+ec_+-z_-ec_- \approx 0,
\label{electroneut1}
\end{equation}
where $z_+$ and $z_-$ are defined as the absolute values of the charge of the cation and anion, respectively.  We will derive the ambipolar diffusivity, which assumes we have a binary $z:z$ electrolyte.  

For porous electrodes, we also need to account for the porosity of the medium.  The porosity affects the interfacial area between volumes of the porous electrode.  It also affects the concentration of a given volume of the electrode.  Accounting for porosity, Equations (\ref{genconseqn1}) and (\ref{gencstflux1}) become
\begin{equation}
\epsilon\frac{\partial c_i}{\partial t} = -\nabla\cdot\textbf{N}_i - R_i
\label{genconseqnporos1}
\end{equation}
and
\begin{equation}
\textbf{N}_i = -\epsilon M_ic_i\nabla\mu_i,
\label{gencstfluxporos1}
\end{equation}
where $\epsilon$ is the porosity, which is the volume of electrolyte per volume of the electrode.  This value may change with position, but this derivation assumes porosity is constant with respect to time.  With this assumption, the Nernst-Planck Equation can be defined for the positive and negative species in the electrolyte.  This yields the cation and anion fluxes,
\begin{eqnarray}
\textbf{N}_+ &=& -\epsilon D_{chem,+}\nabla c_+ - \epsilon \frac{z_+e}{k_BT}D_{+}c_+\nabla\phi \label{cationNP1},\mbox{ and} \\
\textbf{N}_- &=& -\epsilon D_{chem,-}\nabla c_- +\epsilon \frac{z_-e}{k_BT}D_{-}c_-\nabla\phi. \label{anionNP1}
\end{eqnarray}
Next, the flux equations for the cation and anion in Equations (\ref{cationNP1}) and (\ref{anionNP1}) are inserted into Equation (\ref{genconseqnporos1}) and combined with the electroneutrality assumption in Equation (\ref{electroneut1}) to eliminate the potential.  The mass conservation equation is
\begin{eqnarray}
\epsilon\frac{\partial c}{\partial t} &=& \nabla\cdot\left(\epsilon D_{amb}\nabla c\right) - \nabla\cdot\left(\left(\frac{t_+-t_-}{2}\right)\textbf{i}\right)- \nonumber \\
&&\left(\frac{z_+R_+}{2}+\frac{z_-R_-}{2}\right),
\label{masscons1}
\end{eqnarray}
where $t_+$ and $t_-$ are the cation and anion transference numbers, respectively, and $D_{amb}$ is the ambipolar diffusivity.  These values are defined as
\begin{equation}
t_\pm \equiv \frac{z_\pm D_\pm}{z_+D_++z_-D_-},
\label{transference} 
\end{equation}
and
\begin{equation}
D_{amb} \equiv \frac{z_+D_+D_{chem,-}+z_-D_-D_{chem,+}}{z_+D_++z_-D_-}. \label{ambipolar}
\end{equation}
In equation (\ref{masscons1}), $\textbf{i}$ is the current density in the electrolyte, which is given by the sum of the cation and anion fluxes multiplied by their charge, 
\begin{equation}
\textbf{i} = ez_+\textbf{N}_+-ez_-\textbf{N}_-.
\label{currdensdef1}
\end{equation}
Furthermore, the concentration $c$, using the electroneutrality assumption, is defined as
\begin{equation}
c \equiv z_+c_+ = z_-c_-.
\label{concdef1}
\end{equation}
Next, it is necessary to relate the charge conservation to the mass conservation to simplify Equation (\ref{masscons1}).

The electroneutrality approximation puts a restriction on the charge accumulation in the electrolyte.  Since the cations and anions must balance, the divergence of the current density must balance with the ions being produced/consumed via Faradaic reaction in the volume. To determine the charge balance in some volume of the electrode, we begin with the current density as given by Equation (\ref{currdensdef1}).  Simplifying this expression and combining it with the definition of $c$ based on the electroneutrality assumption, the current density is
\begin{eqnarray}
\textbf{i} &=& -e\left(D_{chem,+}-D_{chem,-}\right)\epsilon\nabla c - \nonumber \\
&& \frac{e^2}{k_BT}\left(z_+D_++z_-D_-\right)\epsilon c\nabla\phi.
\label{currdensdef2}
\end{eqnarray}
The divergence of the current density gives the accumulation of charge within a given volume.  As stated above, this value must equal the charge produced or consumed by the reactions within the given volume, therefore
\begin{equation}
ez_+R_+-ez_-R_-=ea_{p,+}j_{in,+}-ea_{p,-}j_{in,-}= -\nabla\cdot\textbf{i},
\label{chargecons1}
\end{equation}
where $a_{p,i}$ is the area per unit volume of the active intercalation particles, and $j_{in,i}$ is the flux into the particles due to Faradaic reactions of species $i$.  For the remainder of the derivation, the term $a_pj_{in}$ will imply the sum of the reaction rates of the species.  Substituting this expression into Equation (\ref{masscons1}) and using the definition $t_++t_-=1$, the conservation equation is
\begin{equation}
\epsilon\frac{\partial c}{\partial t} = \nabla\cdot\left(\epsilon D_{amb}\nabla c\right)+\nabla\cdot\left(\frac{\left(1-t_+\right)\textbf{i}}{e}\right).
\label{masschargecons1}
\end{equation}
Substituting Equation (\ref{chargecons1}) into Equation (\ref{masschargecons1}), the familiar Porous Electrode Theory equation,
\begin{equation}
\epsilon\frac{\partial c}{\partial t} + a_pj_{in} = \nabla\cdot\left(\epsilon D_{amb}\nabla c\right) -{\nabla}\cdot\left(\frac{t_+ \textbf{i}}{e}\right),
\label{PETeqn1}
\end{equation}
is derived.  Since the potential was eliminated in the ambipolar derivation, and the potential gradient is dependent on the current density via Equation (\ref{currdensdef2}), Equations (\ref{currdensdef2}) and (\ref{chargecons1}) can be used to formulate an expression for the local electrolyte potential,
\begin{eqnarray}
a_pj_{in} &=& \nabla\cdot\left[\left(D_{chem,+}-D_{chem,-}\right)\epsilon\nabla c+ \right.
\nonumber \\
&& \left.\frac{e^2}{k_BT}\left(z_+D_++z_-D_-\right)\epsilon c\nabla\phi\right].
\label{PETeqn2}
\end{eqnarray}
Finally, an expression for $j_{in}$ is required to complete the set of equations.  This can be modeled via the Butler-Volmer Equation.  

For phase transforming materials, the activity of the atoms and energy of the transition state can have a dramatic effect on the reaction rate.  To account for this, a modified form of the Butler-Volmer Equation, which accounts for the energy of the transition state, will be derived.

\subsubsection{Faradaic Reaction Kinetics}

The reader is referred to Bazant~\cite{acr2012,10.626} for detailed, pedagogical derivations of Faradaic reaction rates in concentrated solutions and solids, generalizing both the phenomenological Butler-Volmer equation~\cite{bockris_book} and the microscopic Marcus theory of charge transfer~\cite{kuznetsov_book,bard_book,marcus1993}. Here we summarize the basic derivation and focus applications to the case of lithium intercalation in a solid solution.

In the most general Faradaic reaction,  there are $n$ electrons transferred from the electrode to the oxidized state O to produce the reduced state R:
\[
\mbox{O} + ne^- \rightleftharpoons \mbox{R}.
\]
Typically, one electron transfer is favored ~\cite{kuznetsov_book,bard_book,bockris_book}, but for now let us keep the derivation as general as possible.  The reaction goes through a transition state, which involves solvent reorganization and charge transfer. The net reaction rate, $R_{net}$, is the sum of the forward and reverse reaction rates,
\begin{equation}
R_{net} = k\left[\exp\left(-\frac{\mu_{\ddag}^{ex}-\mu_1}{k_BT}\right) - 
\exp\left(-\frac{\mu_{\ddag}^{ex}-\mu_2}{k_BT}\right)\right].
\label{genrxn1}
\end{equation}
Once again, for an isothermal process (which is reasonable at the microscopic scale) the concentration of the transition state is constant and can be factored into the rate constant.

It is first necessary to postulate forms of the {\it electro-}chemical potentials in the generic Faradaic reaction above.  Here it is assumed that both the oxidant and reductant are charged species, and that the electron is at a potential $\phi_M$, which is the potential of the metallic electron-conducting phase (e.g. carbon black). The electrochemical potentials of the oxidant and reductant are broken into chemical and electrostatic contributions as follows:
\begin{equation}
\mu_O = k_BT\ln a_O + eq_O\phi -ne\phi_M + E_O 
\label{oxpostcpot}
\end{equation}
and
\begin{equation}
\mu_R = k_BT\ln a_R + eq_R\phi + E_R,
\label{redpostcpot}
\end{equation}
where $E_O$ and $E_R$ are the reference energies of the oxidant and reductant, respectively.  The excess chemical potential of the transition state is assumed to consist of an activity coefficient contribution and some linear combination of the potentials of the oxidant and reductant,
\begin{equation}
\mu_{\ddag}^{ex} = k_BT\ln\gamma_{\ddag} + \alpha eq_R\phi + (1-\alpha)e\left(q_O\phi-n\phi_M\right) +E_{\ddag},
\label{tspostcpot}
\end{equation}
where $\alpha$, also known as the transfer coefficient, denotes the symmetry of the transition state.  This value is typically between 0 and 1.  Charge conservation in the reaction is given by
\begin{equation}
q_O+n=q_R
\label{chargeconrxn1}
\end{equation}
At equilibrium, $\mu_O=\mu_R$, and the Nernst potential,
\begin{equation}
\Delta\phi_{eq} = V^o + \frac{k_BT}{ne}\ln\left(\frac{a_O}{a_R}\right),
\label{nernsteqn1}
\end{equation}
is obtained, where $V^o=\left(E_O-E_R\right)/ne$.  Equations (\ref{oxpostcpot}), (\ref{redpostcpot}), and (\ref{tspostcpot}) can be substituted directly into the generation reaction rate, (\ref{genrxn1}), to obtain
\begin{eqnarray}
R &=& \frac{k_o}{\gamma_{\ddag}}\left[a_O\exp\left(\tilde{E}_O-\tilde{E}_{\ddag}\right)\exp
\left(-\alpha n\Delta\tilde{\phi}\right) - \right. \nonumber \\
&&\left.a_R\exp\left(\tilde{E}_R-\tilde{E}_{\ddag}\right)
\exp\left(\left(1-\alpha\right)n\Delta\tilde{\phi}\right)\right],
\label{rxnsub1}
\end{eqnarray}
where the energy is scaled by the thermal energy and the voltage is scaled by the thermal voltage.  Next, the definition of overpotential is substituted into Equation (\ref{rxnsub1}).  The overpotential is defined as
\begin{equation}
\eta \equiv \Delta\phi -\Delta\phi_{eq}.
\label{overpotential}
\end{equation}
Combining the definition of the overpotential with the Nernst equation and substituting into Equation (\ref{rxnsub1}), after simplifying we obtain the Modified Butler-Volmer Equation,
\begin{equation}
ej_{in} = i_o\left[\exp\left(-\alpha\tilde{\eta}\right)-
\exp\left(\left(1-\alpha\right)\tilde{\eta}\right)\right],
\label{modbv1}
\end{equation}
where $i_o$, the exchange current density, is defined as
\begin{equation}
i_o = \frac{nek^o\left(a_O\right)^{(1-\alpha)n}\left(a_R\right)^{\alpha n}}{\gamma_{\ddag}},
\label{excurrdensdef}
\end{equation}
and $k^o$, the rate constant, is given by
\begin{equation}
k^o = k_o\exp\left(\alpha n \tilde{E}_R+\left(1-\alpha\right)n\tilde{E}_O-\tilde{E}_\ddag\right)
\label{excurrrateconst1}
\end{equation}
The main difference is that the overpotential and exchange current are defined in terms of the activities of the oxidized, reduced and transition states, each of which can be expressed variationally in terms of the total free energy functional of the system (below). 

Using the  Butler-Volmer Equation, the value of $j_{in}$ (the flux into the particles due to Faradaic reactions) can be modeled.  The overpotential is calculated via the definition given in Equation (\ref{overpotential}), and the equilibrium potential is given by the Nernst Equation, where the activity of the surface of the active material is used.

\subsubsection{Potential Drop in the Conducting Solid Phase}

The reaction rate at the surface of the particles is dependent on the potential of the electron as well as the potential of lithium in the electrolyte.  This is expressed as $\Delta\phi$, which contributes to the overpotential in Equation (\ref{overpotential}). The potential difference is the difference between the electron and lithium-ion potential,
\[
\Delta\phi = \phi_M - \phi,
\]
where $\phi_M$ is the potential of the metallic electron-conducting phase  (e.g. carbon black) phase and $\phi$ is the potential of the electrolyte.  The potential of the electrolyte is determined by the charge conservation equation in Equation (\ref{chargecons1}).  To determine the potential drop in the conducting phase, we use current conservation which occurs throughout the entire electrode, given by 
\begin{equation}
\textbf{i} + \textbf{i}_M = I/A_{sep},
\label{currcons1}
\end{equation}
where $\textbf{i}_M$ is the current density in the carbon black phase.  For constant current discharge, the relation between the local reaction rate and the divergence of the current density in the conducting phase is
\begin{equation}
ea_{p,+}j_{in,+}-ea_{p,-}j_{in,-} = \nabla\cdot\textbf{i}_M.
\label{currcons2}
\end{equation}
The current density in the conducting phase can be expressed using Ohm's Law.  For a given conductivity of the conducting phase, the current density is
\begin{equation}
\textbf{i}_M = -\sigma_m\nabla\phi.
\label{currdensmetal1}
\end{equation}
The conductivity of the conducting phase can be modeled or fit to experiment based on porosity, the loading percent of the carbon black, and/or the lithium concentration in the solid,
\[
\sigma_m = \sigma_m\left(\tilde{c}_s,L_p,\epsilon\right).
\]
As lithium concentration increases in the particles, there are more electrons available for conduction.  These are a few of the cell properties that can have a large impact on the conductivity of the solid matrix in the porous electrode.

\subsubsection{Diffusion in the Solid}

\begin{figure}[htp]
\centering	
	\includegraphics[width=3in,keepaspectratio]{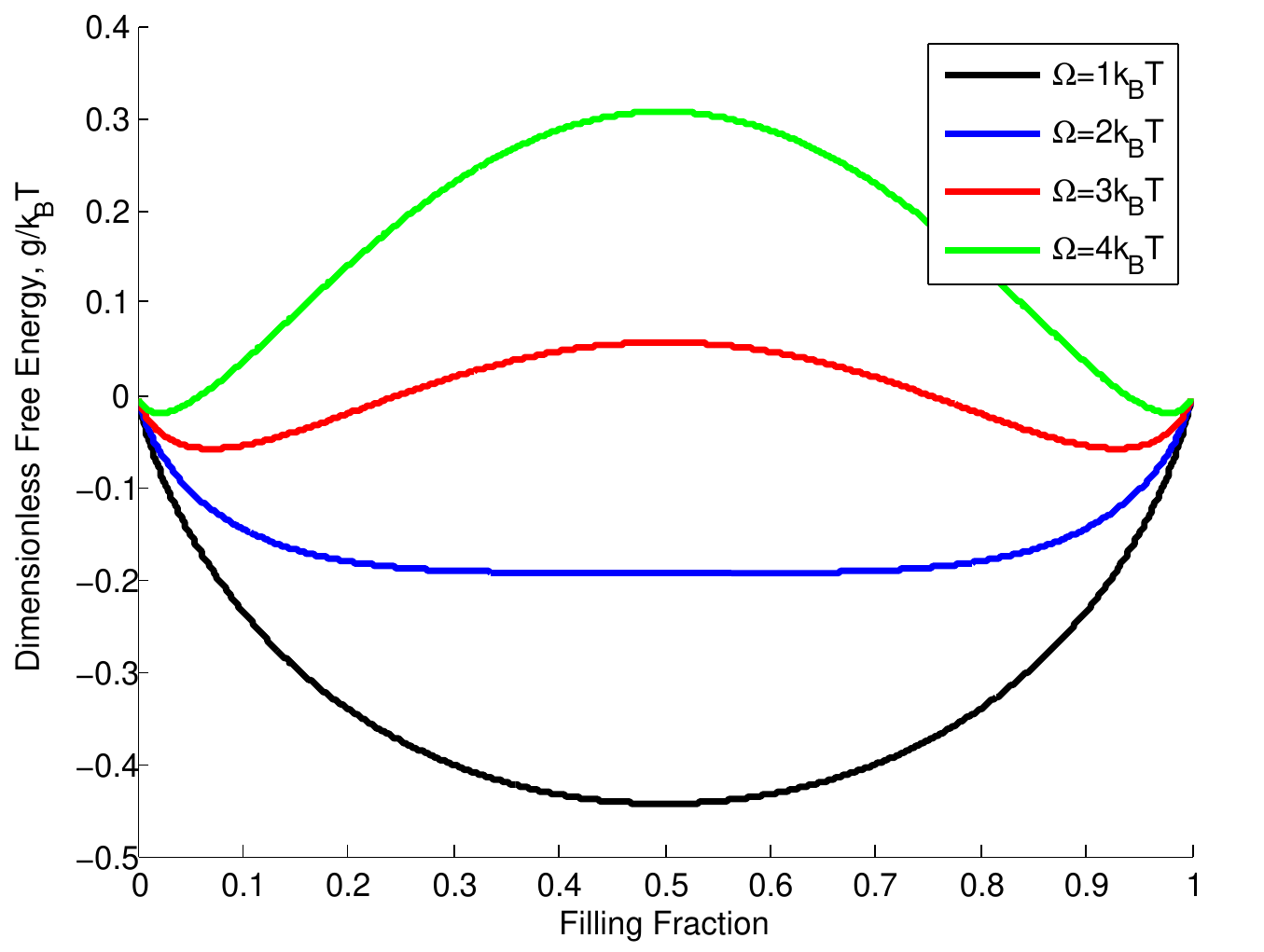}
\caption{\textbf{Regular solution model for the free energy of a homogeneous mixture}.  This figure shows the effect of the regular solution parameter $\Omega$ (mean pair interaction energy)  and temperature $T$ on the free energy versus composition $c$ of a regular solution of atoms and vacancies on a lattice.  For $\Omega<2k_BT$, there is a single minimum.  For $\Omega>2k_BT$, there are two minima.  This produces phase separation, as the system is unstable with respect to infinitesimal perturbations near the spinodal concentration, which is where the curvature of the free energy changes.}
\label{rsfreeenergy1}
\end{figure}

Proper handling of diffusion in the solid particles requires the use of concentrated solution theory.  Diffusion inside solids is often non-linear, and diffusivities vary with local concentration due to finite volume and other interactions inside the solid.  The first section on concentrated solution theory laid the groundwork for proper modeling of diffusion inside the solid.  Here, we begin with the flux defined in Equation (\ref{gencstflux1}),
\[
\textbf{N}_i = -M_ic_i\nabla\mu_i,
\]
where $\textbf{N}_i$ is the flux of species $i$, $M_i$ is the mobility, $c_i$ is the concentration, and $\mu_i$ is the chemical potential.  With no sink or source terms inside the particles, the mass conservation equation from Equation (\ref{genconseqn1}) is
\begin{equation}
\frac{\partial c_i}{\partial t} = -\nabla\cdot\textbf{N}_i.
\label{consdiffeqn1}
\end{equation}
There are many different models which can be used for the chemical potential.  For solid diffusion, one model that is typically used is the regular solution model, which incorporates entropic and enthalpic effects.\cite{hildebrand1929,cahn1958,kom}.  The regular solution model free energy is
\begin{equation}
g = k_BT\left[\tilde{c}\ln\tilde{c}_s +\left(1-\tilde{c}_s\right)\ln\left(1-\tilde{c}_s\right)\right] + \Omega\tilde{c}_s\left(1-\tilde{c}_s\right),
\label{regsolnfree1}
\end{equation}
where $\tilde{c}_s$ is the dimensionless solid concentration ($\tilde{c}_s = c_s/c_{s,max}$).  Figure \ref{rsfreeenergy1} demonstrates the effect of the regular solution parameter (i.e. the pairwise interaction) on the free energy of the system.  The model is capable of capturing the physics of homogeneous and phase separating systems.

Homogeneous particles demonstrate solid solution behavior, as all filling fractions are accessible.  This behavior is typically indicated by a monotonically decreasing open circuit voltage curve.  In terms of the regular solution model, a material that demonstrates solid solution behavior has a regular solution parameter of less than $2k_BT$, that is $\Omega<2k_BT$.  This is related to the free energy curve.  When $\Omega\leq 2k_BT$, there is a single minimum in the free energy curve over the range of concentrations.  However, for $\Omega>2k_BT$, there are two minima, resulting in phase separation and a common tangent, which corresponds to changing fractions of each phase.

The common tangent construction arises from the fact that phases in equilibrium have the same chemical potential (i.e. slope).  The chemical potential of the regular solution model is
\begin{equation}
\mu = \frac{\partial g_i}{\partial c_{s,i}} = k_BT\ln\left(\frac{\tilde{c}_s}{1-\tilde{c}_s}\right)+
\Omega\left(1-2\tilde{c}_s\right).
\label{regsoln1}
\end{equation}
To obtain an analogous equation to Fick's First Law, Equation (\ref{gencstflux1}) can be expressed as
\begin{equation}
\textbf{N}_i = -D_o\left(1-\tilde{c}_s\right)
\left(1+\frac{\partial\ln\gamma_i}{\partial\ln c_{s,i}}\right)\nabla c_{s,i} = - D_{chem} \nabla c_{s,i},
\label{cstflux2}
\end{equation}
where $D_o$ is the diffusivity of species $i$ in the solid in the infinitely dilute limit and $D_{chem}$ is the chemical diffusivity in a concentrated solution.  It is important to note that $D_o$ can still be a function of concentration.  The regular solution model in Equation (\ref{regsoln1}) can be substituted into Equation (\ref{cstflux2}) using the definition of the chemical potential, $\mu=k_BT\ln(c\gamma)$, to obtain the chemical diffusivity,
\begin{equation}
D_{chem} = D_o\left(1-2\tilde{\Omega}\tilde{c}_s+2\tilde{\Omega}\tilde{c}_s^2\right),
\label{rseffectivediff1}
\end{equation}
where $\tilde{\Omega} = \Omega/k_BT$, the dimensionless interaction energy. When the interaction parameter, $\Omega$, is zero, the dilute limit diffusivity (Fick's Law) is recovered.  The mass conservation equation using the effective diffusivity is
\begin{equation}
\frac{\partial c_s}{\partial t} = \nabla\cdot\left(D_{chem}\nabla c_s\right).
\label{soliddiff1}
\end{equation}
Phase separating materials (e.g. LiFePO$_4$) can be described by the Cahn-Hilliard free energy functional,\cite{cahn1958} 
\begin{equation}
G[\tilde{c}(x)] = \int_V\left[\rho_s\overline{g}(\tilde{c})+ \frac{1}{2}\kappa\left(\nabla \tilde{c}\right)^2\right]dV + \int_A \gamma_s\left(\tilde{c}\right) da,
\label{cahnhilliardFE1}
\end{equation}
where $\overline{g}\left(\tilde{c}\right)$ is the homogeneous bulk free energy, $\rho_s$ is the site density, $\kappa$ is the gradient energy (generally, a tensor for an anisotropic crystal), with units of energy per length, and $\gamma_s\left(\tilde{c}\right)$ is the surface tension, which is integrated over the surface area $A$ to obtain the total surface energy.  The ``gradient penalty" (second term) can be viewed as the first correction to the free energy for heterogeneous composition, in a perturbation expansion about the homogeneous state. When phase separation occurs, the gradient penalty controls the structure and energy of the phase boundary between stable phases (near the minima of $\overline{g}(\tilde{c})$). For example, balancing terms in (\ref{cahnhilliardFE1}) in the case of the regular solution model, the phase boundary width scales as $\lambda_i \approx \sqrt{\kappa/\Omega}$, and the interphasial tension as $\gamma_i \approx \sqrt{\kappa \Omega} \rho_s$ ~\cite{kom,cahn1958,burch2009}. 

More complicated phase-field models of the total free energy can also be used in our general porous electrode theory. For example, elastic coherency strain can be included with additional bulk stress-strain terms~\cite{garcia2004,vanderven2009,cogswell2012}, as described below. It is also possible to account for diffuse charge and double layers by incorporating electrostatic energy in the total free energy functional~\cite{garcia2004,guyer2004a,guyer2004b,bazant2011,acr2012}, although we neglect such effects here and assume quasi-neutrality in the electrolyte and active solid particles.

Once the total free energy functional is defined, the chemical potential of a given species is defined by the Euler-Lagrange variational derivative  with respect to concentration, which is the continuum equivalent of the change in free energy to ``add an atom" to the system.  The chemical potential per site is thus 
\begin{equation}
\mu = \frac{1}{\rho_s}\frac{\delta G}{\delta \tilde{c}} = \overline{\mu}\left(\tilde{c}\right) - \nabla\cdot\left(\frac{\kappa}{\rho_s}\nabla \tilde{c}\right),
\label{cahnhilliardCPot1}
\end{equation}
where $\overline{\mu}$ is the homogeneous chemical potential. 
Using Equation (\ref{gencstflux1}), the flux is based on the gradient of the chemical potential, and the conservation equation is
\begin{equation}
\frac{\partial c}{\partial t} = \nabla\cdot\left(Mc\nabla\mu\right).
\label{cstcons5}
\end{equation}
For typical second-order diffusion equations, the boundary condition relates the normal flux to the reaction rate of each species.  
When the Cahn-Hilliard chemical potential is used in Equation (\ref{cahnhilliardCPot1}), however, the conservation equation contains a fourth derivative of concentration, requiring the use of another boundary condition.  The calculus of variations provides the additional ``variational boundary condition",
\begin{equation}
\hat{n}\cdot\kappa \nabla c_i = \frac{\partial \gamma_s}{\partial c_i}   \label{eq:wet}
\end{equation}
which ensures continuity of the chemical potential~\cite{burch2009} and controls surface wetting and nucleation~\cite{bai2011}.

The choice of the gradient and divergence operators is dependent upon the selected geometry of the particles.  To complete the modeling of the particles, we impose two flux conditions: one at the surface and the other at the interior of the particle.  For example, consider a spherical particle with a radius of 1.  The boundary conditions are
\begin{equation}
\left.\frac{\partial c}{\partial r}\right|_{r=0} = 0
\label{solidbc1}
\end{equation}
and
\begin{equation}
-D_s\left.\frac{\partial c}{\partial r}\right|_{r=1} = j_{in},
\label{solidbc2}
\end{equation}
where $D_s$ is the solid diffusivity (can be a function of concentration).  These equations demonstrate the symmetry condition at the interior of the particle, and the relation to the reaction rate at the surface of the particle, which comes from the modified Butler-Volmer Equation.

\subsubsection{Modeling the Equilibrium Potential}

\begin{figure}[htp]
\centering	
	\includegraphics[width=3in,keepaspectratio]{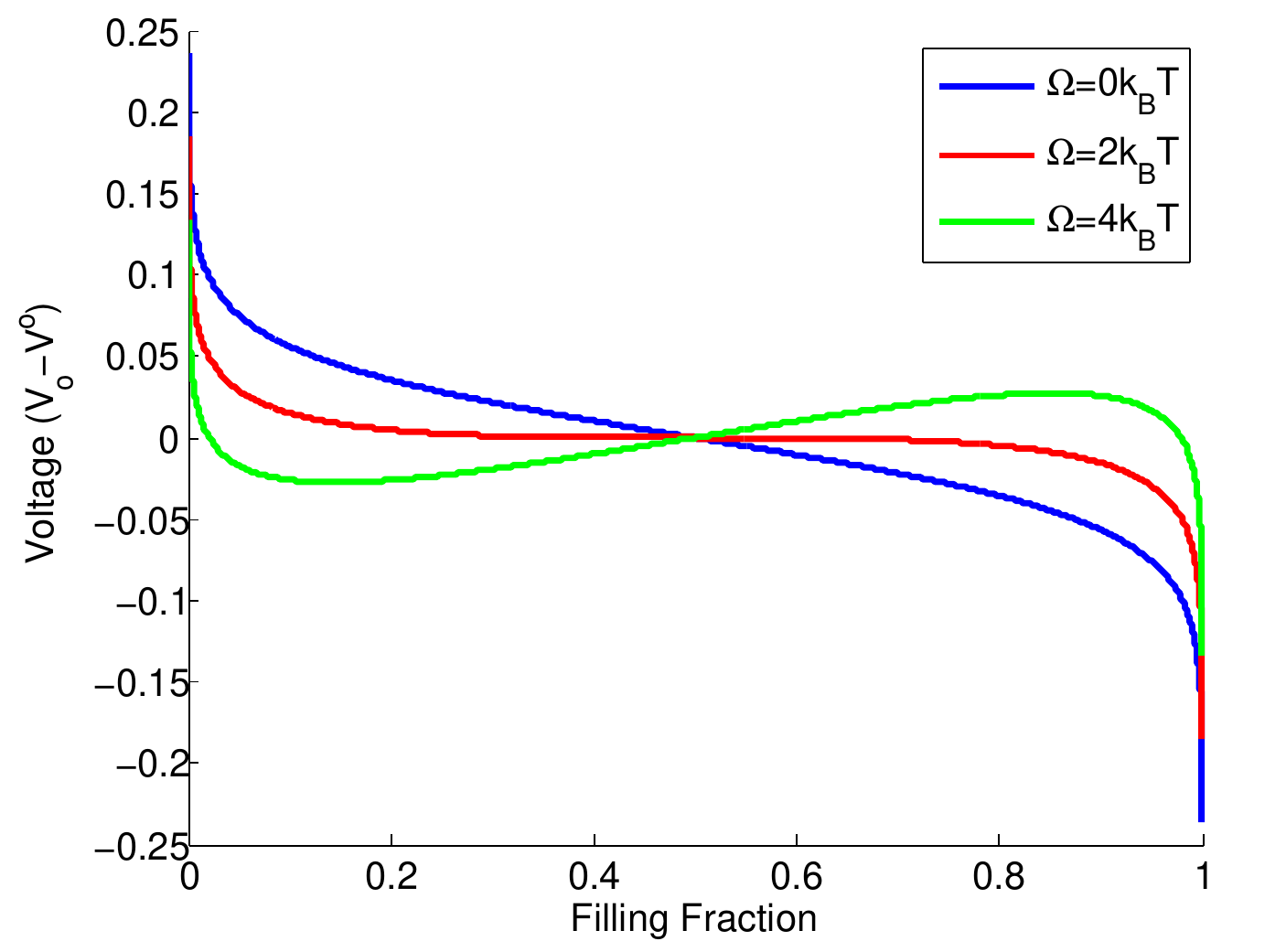}
\caption{\textbf{Open circuit potential for different regular solution parameter values}.  The battery voltage is the change in free energy per electron transferred.  In this model, the homogeneous voltage curve is non-monotonic when the system has a tendency for phase separation. }
\label{ocpcomp1}
\end{figure}

To complete the model, a form of the open circuit potential (OCP) is required.  While traditional battery models fit the OCP to discharge data, the OCP is actually a function of the thermodynamics of the material.  The OCP can be modeled using the Nernst Equation given in Equation (\ref{nernsteqn1}), 
\[
\Delta\phi_{eq} = V^o-\frac{k_BT}{ne}\ln\left(\frac{a_R}{a_O}\right),
\]
where $V^o$ is the standard potential.  Typically, we take lithium metal as the reference potential for the anode and cathode materials.  For the cathode material, this allows us to treat the activity of the oxidant as a constant.  Let's again consider the regular solution model.  Using the definition for chemical potential, $\mu\equiv k_BT\ln a$, we substitute in our regular solution chemical potential to get
\begin{equation}
\Delta\phi_{eq} = V^o-\frac{k_BT}{e}\ln\left(\frac{\tilde{c}_s}{1-\tilde{c}_s}\right) - \frac{\Omega}{e}\left(1-\tilde{c}_s\right).
\label{regsolnocplm1}
\end{equation}
Figure \ref{ocpcomp1} shows open circuit potential curves for different regular solution parameter values.  For $\Omega>2k_BT$, the system is phase separating.  This corresponds to a non-monotonic voltage diagram.

Since the reaction occurs at the surface, and the concentration inside the solid is not necessarily uniform, then surface concentration determines the local OCP.  This in turn affects the overpotential and the reaction rate.  Larger overpotentials are required when the solid has a slow diffusivity.  As lithium builds up at the surface of the particle, a higher overpotential is required to drive the intercalation reaction.

\subsection{Non-Dimensionalization and Scaling}

In this section, the equations are non-dimensionalized for the full three dimensional case.  Here we assume the anode is lithium metal with fast kinetics.  This allows us to model the separator and cathode.  This non-dimensionalization can easily be expanded to model the anode as well.  The electrode is assumed to have a constant cross sectional area, which is typical in rolled electrodes where the area of the separator is much larger than the electrode thickness.  The total current is the sum of the fluxes into the particles in the electrode.  This is represented by the integral equation
\begin{equation}
I = \int_{A_s}ej_{in}dA_s = \int_{V_s}ea_pj_{in}dV_s,
\label{totalcurr1}
\end{equation}
where $a_p$ is the area per volume of the particles.  The solid volume, $V_s$, can be expressed as $\left(1-\epsilon\right)L_pV$, where $\epsilon$ is the porosity, $L_p$ is the volume fraction of active material, and $V$ is the volume of the cell.  Scaling the time by the diffusive time (in the dilute limit), $t_d = L^2/D_{amb,o}$, and the charge by the capacity of the entire electrode, the dimensionless current is
\begin{equation}
\tilde{I}=\frac{It_d}{e\left(1-\epsilon\right)P_LVc_{s,max}} = \int_{\tilde{V}}\tilde{j_{in}}d\tilde{V},
\label{ndtotalcurr1}
\end{equation}
where the dimensionless reaction flux, $\tilde{j}_{in}$, is defined as
\begin{equation}
\tilde{j}_{in} = \frac{a_pj_{in}t_d}{c_{s,max}}.
\label{ndcurrdens1}
\end{equation}

The non-dimensional current density in the electrolyte is
\begin{equation}
\tilde{\textbf{i}} = -\left(\tilde{D}_{chem,+} -\tilde{D}_{chem,-}\right)\tilde{\nabla}\tilde{c}-\left(z_+\tilde{D}_++z_-\tilde{D}_-\right) \tilde{c}\tilde{\nabla}\tilde{\phi},
\label{ndcurrdensdef2}
\end{equation}
where the dimensionless current density $\tilde{\textbf{i}}$ is defined as
\begin{equation}
\tilde{\textbf{i}} = \frac{t_d\textbf{i}}{Lec_o}.
\label{ndcurrdensdefi}
\end{equation}
The diffusivities in the dimensionless current density equation above are scaled by the dilute limit ambipolar diffusivity.  Similarly, the non-dimensional charge conservation equation becomes
\begin{equation}
\beta\tilde{j}_{in} = -\tilde{\nabla}\cdot\tilde{\textbf{i}},
\label{ndchargecons1}
\end{equation}
where $\beta = V_sc_{s,max} / V_ec_o$ is the ratio of lithium capacity in the solid to initial lithium in the electrolyte.  This parameter is important, as it determines the type of cell.  For $\beta\ll 1$, the system has essentially no storage capability, and the equations are typically used to model capacitors.  At $\beta\approx 1$, the system has comparable storage in the electrolyte and solid.  This is typically seen in pseudocapacitors.  The equations for systems like these typically include a term for double layer charge storage as well.  For $\beta\gg 1$, there is a large storage capacity in the solid, which is typically found in batteries.

Next, a mass balance on the electrolyte and solid are performed.  Equation (\ref{PETeqn1}) is non-dimensionalized for some control volume inside the electrode.  In this control volume, the electrolyte and solid volumes are represented by $V_e$ and $V_s$, respectively.  It is assumed that the electrode has the same properties throughout (e.g. porosity, loading percent, area per volume, etc.).  The dimensionless mass balance is
\begin{equation}
\frac{\partial\tilde{c}}{\partial\tilde{t}}+\beta\tilde{j}_{in}=\tilde{\nabla}
\cdot\left(\tilde{D}_{amb}\tilde{\nabla}\tilde{c}\right)-\tilde{\nabla}\cdot\left(
t_+\tilde{\textbf{i}}\right),
\label{PETndeqn2}
\end{equation}
where the time is scaled by the diffusive time scale, $t_d$, the gradients are scaled by the electrode length, $L$, the diffusivity is scaled by the dilute limit ambipolar diffusivity, $D_{amb,o}$, the electrolyte concentration is scaled by the initial electrolyte concentration, $c_o$, and the current density, $j_{in}$, is scaled as in Equation (\ref{ndcurrdens1}).  

Next, we need to find the dimensionless boundary conditions for the system.  This can be done via integrating the equations over the volume of the cell (in this case the separator and cathode, but this can easily be extended to include the anode).  Integrating Equation (\ref{PETndeqn2}) over the volume yields
\begin{equation}
\int_{\tilde{V}}\left[\frac{\partial\tilde{c}}{\partial\tilde{t}}+\beta\tilde{j}_{in}=
\tilde{\nabla}\cdot\left(\tilde{D}_{amb}\tilde{\nabla}\tilde{c}\right)-\tilde{\nabla}
\cdot\left(t_+\tilde{\textbf{i}}\right)\right]d\tilde{V}.
\label{PETndInt1}
\end{equation}
First, we deal with the left most term.  Given the electroneutrality constraint, this term becomes zero because the amount of anions in the system remains constant.  This assumes no SEI growth.  If SEI growth is modeled, then this term will be related to the time integral of the anion reaction rate.  Integrating the second term, for constant $\beta$, reduces to $\beta\tilde{I}$.  The two terms on the right hand side of the equation facilitate the use of the Fundamental Theorem of Calculus. Simplifying, we obtain
\begin{equation}
\beta\tilde{I} = \left.\left(\tilde{D}_{amb}\tilde{\nabla}\tilde{c} - t_+\tilde{\textbf{i}}\right)\right|_{0}^{1}.
\label{PETndInt2}
\end{equation}
Given the no flux conditions in $\tilde{y}$ and $\tilde{z}$, and the no flux condition at $\tilde{x}=1$, the flux into the separator is
\begin{equation}
\left.-\tilde{D}_{amb}\tilde{\nabla}\tilde{c}\right|_{\tilde{x}=-\tilde{x}_s} = \left(1-t_+\right)\beta\tilde{I}.
\label{PETndInt3}
\end{equation}
This set of dimensionless equations and boundary conditions are used in the simulations presented in the results section.  Table \ref{dimeqntable} lists the equations used in the simulations.

\begin{center}
\begin{table*}[htp]
\begin{tabular}{ b{8.5cm} b{8.5cm} }
\hline
\begin{center}
	Equation 
\end{center} & 
\begin{center}
	Boundary Conditions 
\end{center} \\
\hline
\begin{center}
	$\epsilon\frac{\partial c}{\partial t} + a_pj_{in} = 
	\nabla\cdot\left(\epsilon D_{amb}\nabla c\right) - 
	\nabla\cdot\left(\frac{t_+\textbf{i}}{e}\right)$
\end{center} & 
\begin{center}
	$\left.\textbf{i}\right|_{x=-\delta_s} = I/A_{sep}$ 
\end{center} \\
\begin{center}
	$\textbf{i} = -e\left(D_+-D_-\right)\epsilon\nabla c - 
	\frac{e^2}{k_BT}\left(z_+D_+ + z_-D_-\right)\epsilon c\nabla\phi$ 
\end{center} & 
\begin{center}
\end{center} \\
\begin{center}
	$j_{in} = -\frac{\nabla\cdot\textbf{i}}{ea_p} = 
	i_o\left[\exp\left(-\frac{\alpha 
	e\eta}{k_BT}\right)-\exp\left(\frac{(1-\alpha)e\eta}{k_BT}\right)\right]$
\end{center} & 
\begin{center}
\end{center} \\
\begin{center}
	$i_o = \frac{e\left(k_c^oa_O\right)^{1-\alpha} 
	\left(k_a^oa_R\right)^{\alpha}}{\gamma_{\ddag}}$
\end{center} & 
\begin{center}
\end{center} \\
\begin{center}
	$\eta\equiv \Delta\phi - \Delta\phi_{eq}$
\end{center} & 
\begin{center}
\end{center} \\
\begin{center}
	$\Delta\phi_{eq} = V^o-\frac{k_BT}{ne}\ln\left(\frac{a_R}{a_O}\right)$
\end{center} & 
\begin{center}
\end{center} \\
\begin{center}
	$\frac{\partial c_s}{\partial t} = \nabla\cdot\left(\frac{D_sc_s}{k_BT} \nabla\mu \right)$
\end{center} & 
\begin{center}
	$\left.-\frac{D_sc_s}{k_BT}\frac{\partial \mu}{\partial r}\right|_{r=R} = j_{in}$
\end{center} \\
\begin{center}
\end{center} \\
\end{tabular}
\caption{\textbf{Dimensional set of equations}.  A list of the set of dimensional equations for Modified Porous Electrode Theory.}
\label{dimeqntable}
\end{table*}
\end{center}

\section{Model Results}

To characterize the properties of the model, we will demonstrate some results from the non-dimensional model. Again it is assumed that the anode is lithium metal with fast kinetics, allowing us to model the separator and cathode.  Results for monotonic (i.e. homogeneous) and non-monotonic (i.e. phase separating) open circuit potential profiles for particles demonstrating solid solution behavior will be given for constant current discharge. 

The electrolyte concentration, electrolyte potential, and solid concentration are all coupled via the mass and charge conservation equations listed above.  Solving these equations is often done via Crank-Nicholson and use of the BAND subroutine, which is used to solve the system of equations. \cite{newman_book} Botte \emph{et al.} have reviewed the numerical methods typically used to solve the porous electrode equations. \cite{botte2000}   The system of equations presented in this paper was solved using MATLAB and its \emph{ode15s} differential algebraic equation (DAE) solver.  This code utilizes the backwards differentiation formula (BDF) for time stepping and a dogleg trust-region method for its implicit solution.  The spatial equations were discretized using a finite volume method.  Constant current discharge involves an integral constraint on the system.  This integral constraint makes the system ideal for formulating the system of equations as a DAE.  Formulation of the system of equations as well as some basic numerical methods employed in solving these types of DAE's will be the focus of a future paper. 

These results will highlight the range of physics in the model, which include electrolyte diffusion limited discharge and solid diffusion limited discharge.  These two limitations represent the most common situations in a cell.  Another common limitation is electron conductivity in the solid matrix.  This limitation is often suppressed via increasing the amount of conductive additive used.  Furthermore, some active materials naturally conduct electrons, alleviating this effect.  

The electrolyte diffusion limitation can also be alleviated with proper cell design (i.e. thinner electrode), but this comes at the cost of capacity of the cell.  To demonstrate the effect of electrolyte diffusivity limitations and solid diffusivity limitations, different discharge rates were selected and different solid diffusivities were modeled.  First, we consider the case of homogeneous particles.  Then we demonstrate phase separating particles using the Cahn-Hilliard free energy functional with and without approximated stress effects. 

\subsection{Simulation Values}

The ambipolar diffusivity (given by Equation (\ref{ambipolar})) is taken from literature values for the diffusivity of Li$^+$ and PF$_6^-$ in an EC/EMC non-aqueous electrolyte.  Using literature values for the diffusivities, a value of $1.9\times 10^{-10}$ m$^2$s$^{-1}$ was calculated for $D_{amb,o}$. \cite{capiglia1999,valoen2005}  Suitable cell size parameters were used, including a cross sectional area of 1 cm$^2$, separator thickness of $25 \mu$m, and an electrode length of $50 \mu$m.  A porosity value of 0.4 was used, which is a little larger than typical cell values.  While cell dimensions are typically fixed, the ambipolar diffusivity and porosity values are flexible, and can be varied (within reason) to fit experimental data.

Using these cell dimensions and ambipolar diffusivity, the diffusive time scale for the system is 13.125 seconds.  This value is important, as it affects the non-dimensional total current (which is scaled by the electrode capacity and the diffusive time), the non-dimensional current density, and the non-dimensional exchange current density (i.e. rate constant).  Using this value of the ambipolar  diffusivity, a dimensionless current of $\tilde{I} = 0.00364$ corresponds to approximately a 1C discharge.  The solid diffusivity is incorporated in a dimensionless parameter,
 \begin{equation}
 \delta_d = \frac{L_s^2 D_{amb}}{L^2 D_s }
 \end{equation}
which is the ratio of the diffusion time in the solid ($L_s^2/D_s$) to the diffusion time in the electrolyte ($L^2/D_{amb}$).  This parameter, which is typically typically larger than one, can vary by orders of magnitude for different materials.  Typically, solid diffusivities are unknown, and this parameter needs to be fit to data. 

The rate constant, which directly affects the exchange current density, is another value that is unknown in the system.  The dimensionless value of the exchange current density is scaled to the diffusive time.  It also depends on the average particle size, as this gives the surface area to volume ratio.  For 50 nm particles, using the ambipolar diffusivity above, a dimensionless exchange current density of one corresponds to approximately 1.38 A/m$^2$.  This is a relatively high exchange current density.  For the simulations below, a dimensionless exchange current density of 0.01 is used.  It is important to note that this value must be fit to data, though.

\subsection{Homogeneous Particles}

\begin{figure}[htp]
\centering	
	\includegraphics[width=3in,keepaspectratio]{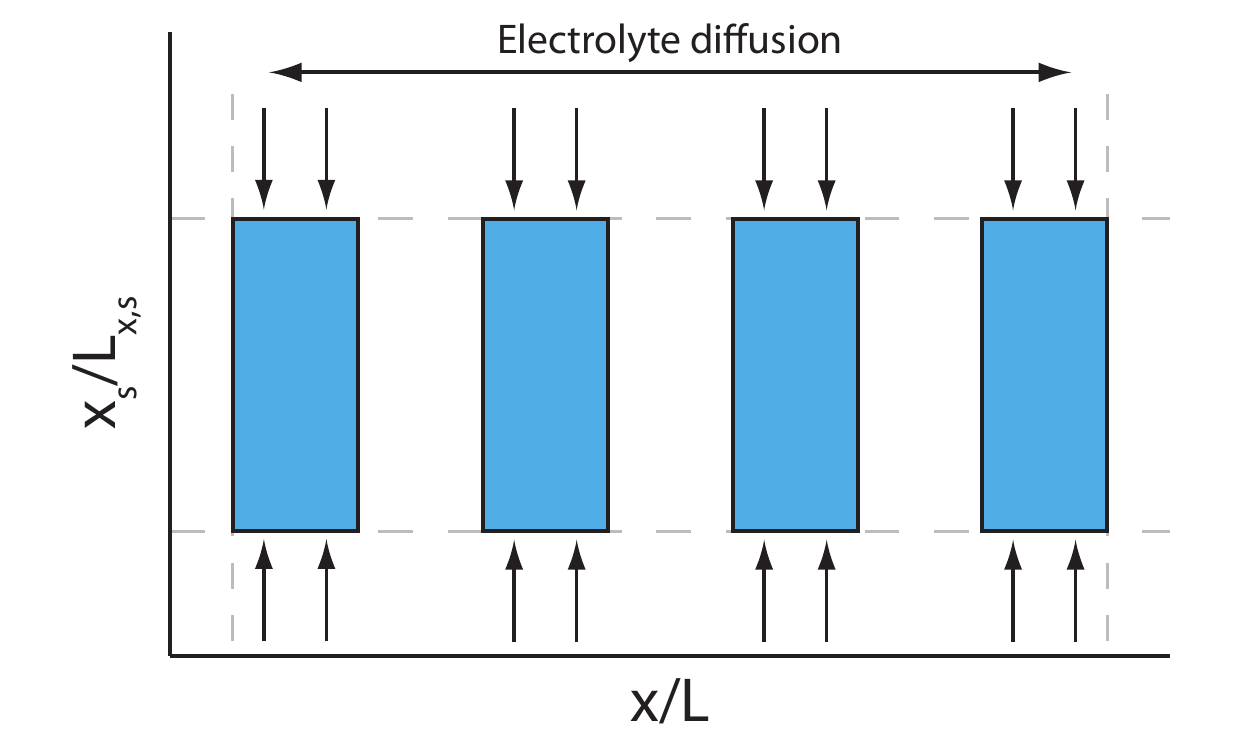}
\caption{\textbf{Plot axes for diffusion-limited solid-solution particles}. This figure shows how the simulation results below are plotted for porous electrodes with isotropic solid solution particles.  The y-axis of the contour plots represent the depth of the particles while the x-axis represents the depth into the electrode.  The particles are modeled in 1D.}
\label{homdiag1}
\end{figure}

\begin{figure}[htp]
\centering	
	\includegraphics[width=3in,keepaspectratio]{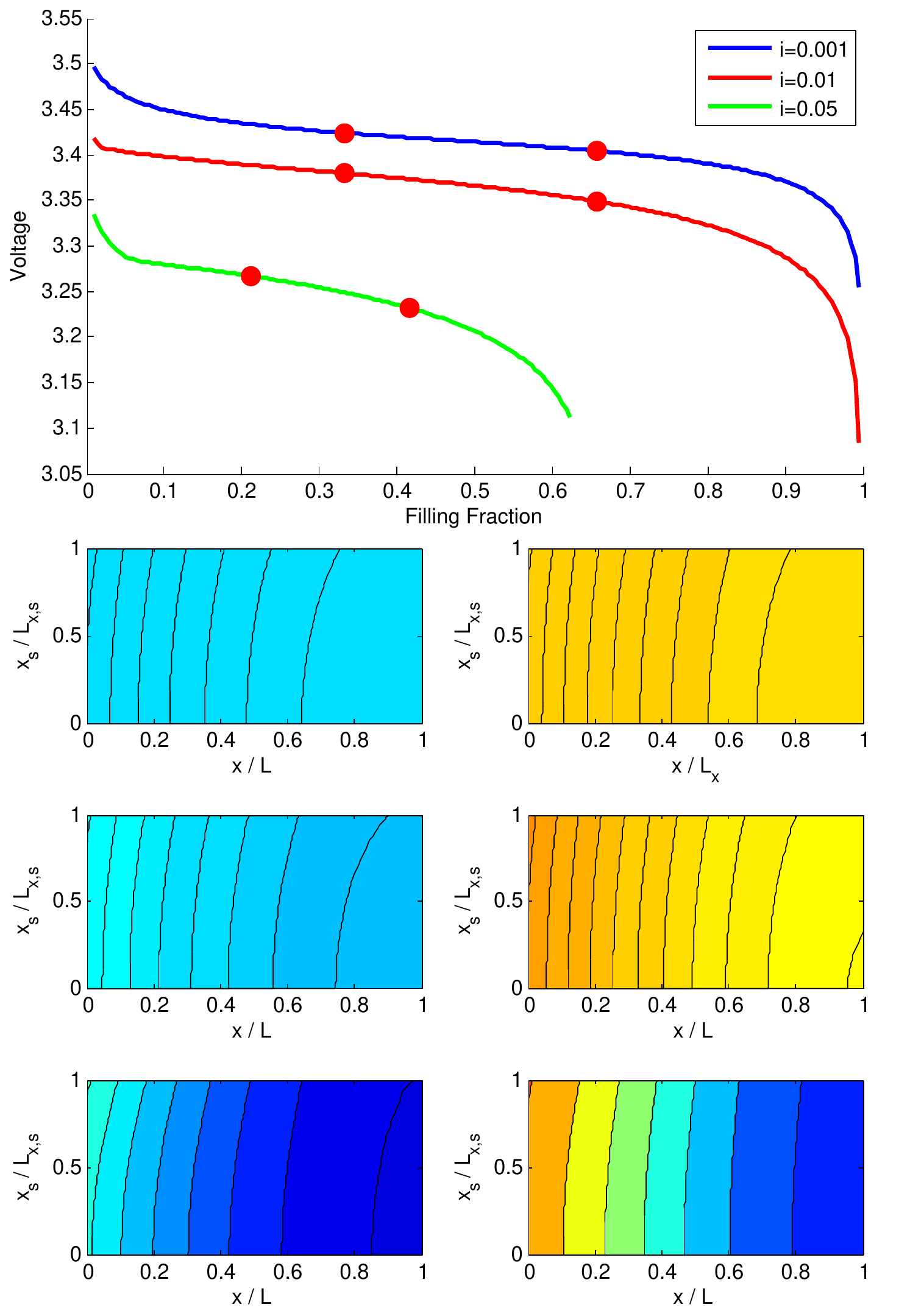}
\caption{\textbf{Effect of current on homogeneous particles}.  This figure demonstrates the effect of different discharge rates on the voltage profile.  The non-dimensional currents correspond to roughly C/3, 3C, and 15C.  The solid diffusion is fast, with $\delta_d=1$.}
\label{homogvarcurr1}
\end{figure}

\begin{figure}[htp]
\centering	
	\includegraphics[width=3in,keepaspectratio]{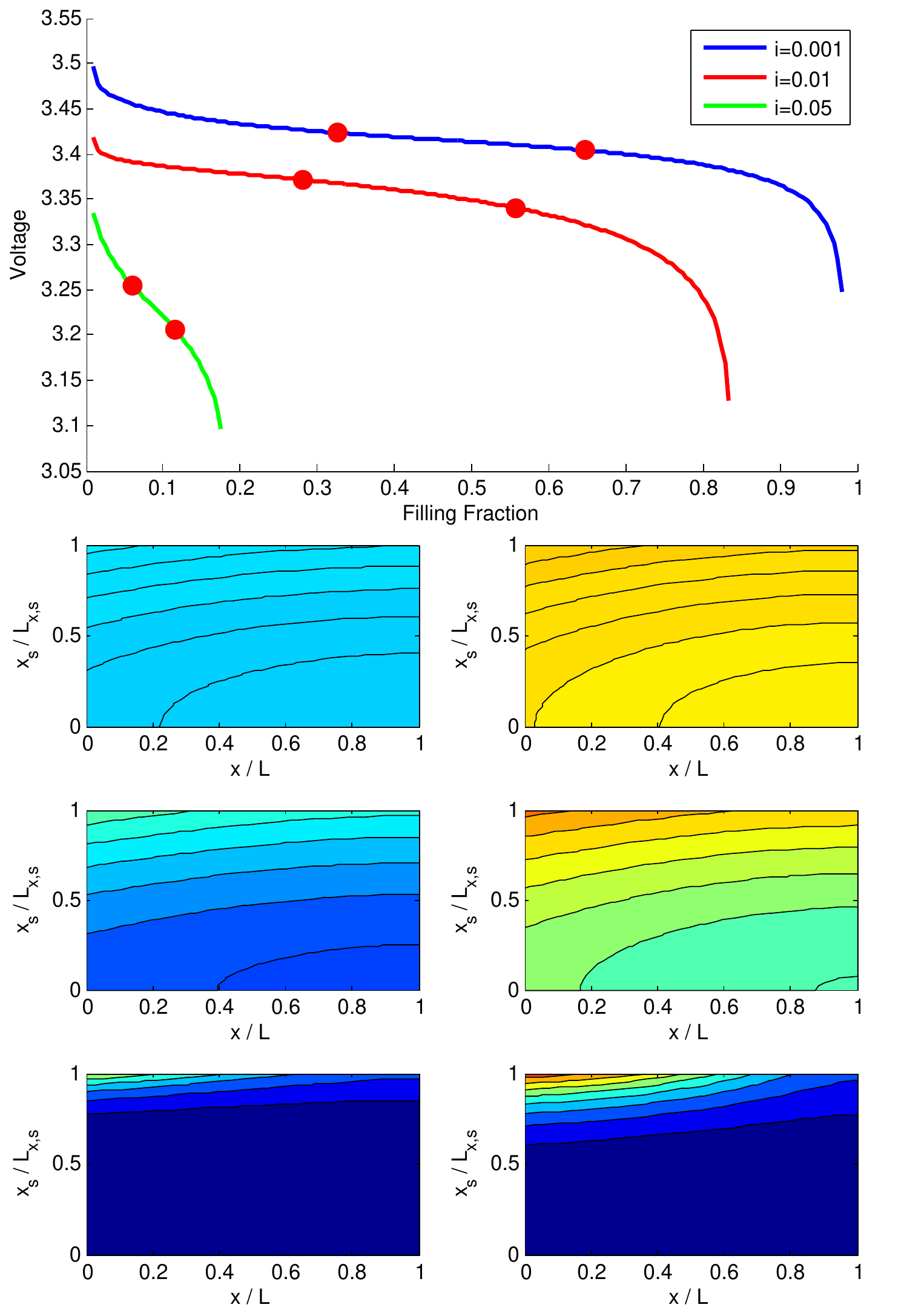}
\caption{\textbf{Effect of current on homogeneous particles with slower solid diffusion}.  This figure demonstrates the effect of different discharge rates on the voltage profile.  The non-dimensional currents correspond to roughly C/3, 3C, and 15C.  The solid diffusion is slower than the electrolyte diffusion ($\delta_d=100$).}
\label{homogvarcurrhighdiff1}
\end{figure}

\begin{figure}[htp]
\centering	
	\includegraphics[width=3in,keepaspectratio]{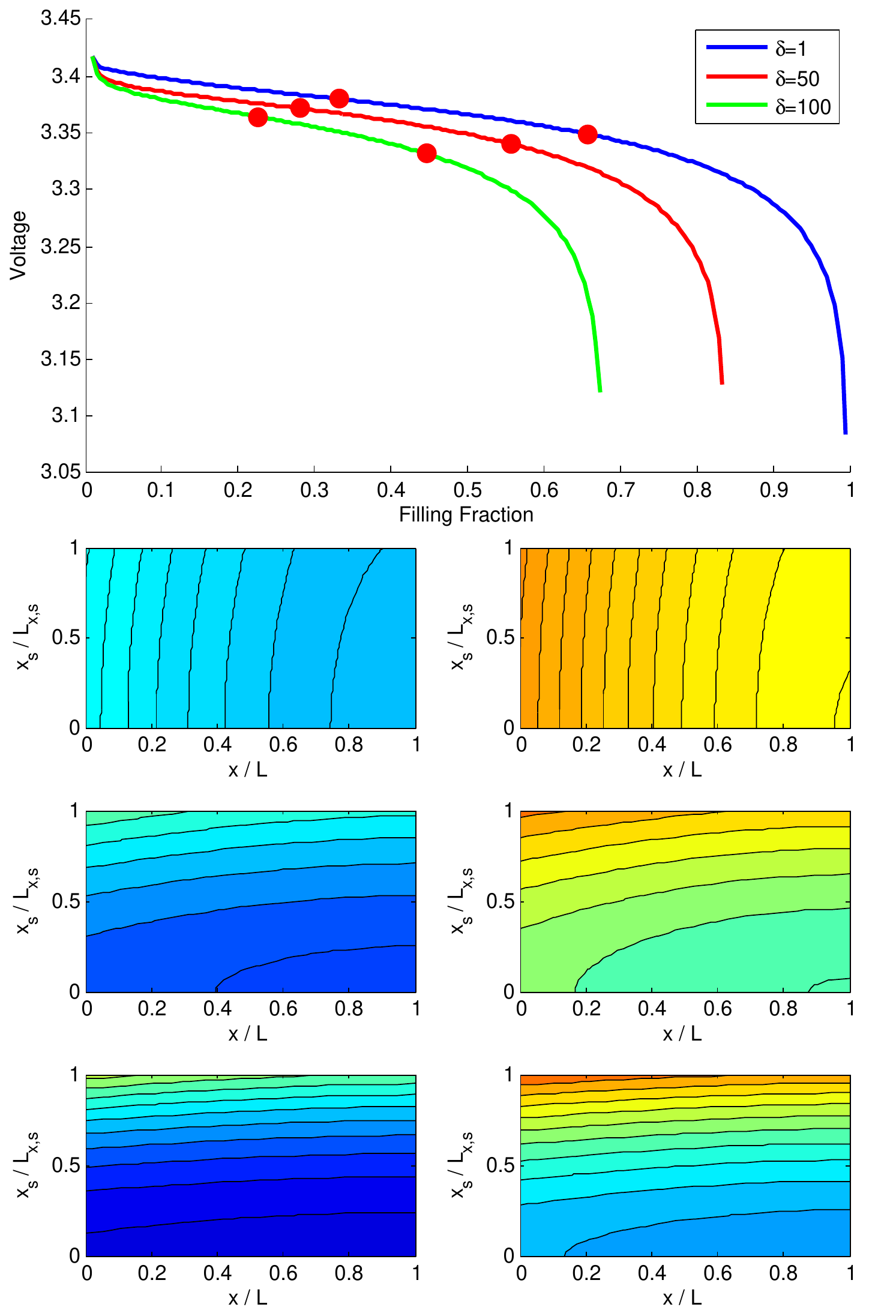}
\caption{\textbf{Effect of solid diffusivity on homogeneous particles}. This figure demonstrates the effect of decreasing solid diffusivity on the voltage profile.  Each of these simulations was run at a dimensionless exchange current density of 0.01 and a dimensionless current of 0.01.}
\label{homogvardiff1}
\end{figure}

Homogeneous particles can access all filling fractions as they are discharged.  Here we consider homogeneous particles using the regular solution model for the open circuit potential and diffusivity inside the solid, as in Equation \ref{rseffectivediff1}.  A value of $\Omega = 1k_BT$ was used.  Figures \ref{homogvarcurr1}, \ref{homogvarcurrhighdiff1}, and \ref{homogvardiff1} demonstrate the effect of various discharge rates and solid diffusivities on the voltage profile.  Each figure contains three different voltage plots.  The red dots on the voltage curves indicate the filling fraction of the solid concentration contours below.  The contour plots are arranged in the same order as the red dots, going from left to right, top to bottom.  Figure \ref{homdiag1} gives the axes for the simulations.  Each particle is modeled in 1D, with the intercalation reaction at the top and diffusion into the bulk of the particle.  The $x_s$ axis is the depth into the particle.

The contour plots give the solid concentration profile of each volume of particles along the length of the electrode.  The y-axis is the depth in the solid particle, with the top ($\tilde{y}=1$) denoting the interface between the particle and the electrolyte.  The x-axis denotes the depth into the electrode, with the left side representing the separator-electrode interface and the right side representing the current collector.  It is important to note that in order for lithium to travel horizontally, it must first diffuse through the solid, undergo a Faradaic reaction to leave the solid, diffuse through the electrolyte, then intercalate into another particle and diffuse.  Therefore sharp concentration gradients in the x-direction are stable, especially for the case of non-monotonic voltage profiles, as is seen in phase separating materials. 

Figure \ref{homogvarcurr1} demonstrates the effect of various discharge rates on the voltage.  At $\tilde{I}=0.001$ (C/3), the discharge is slow and the solid in the electrode fills homogeneously throughout.  As the discharge rate is increased, increased overpotential follows.  Furthermore, gradients in solid concentration down the length of the electrode begin to emerge.  Concentration gradients within the solid are not present because of the high solid diffusivity ($\delta_d=1$, indicating the solid and electrolyte diffusive time scales are the same).  

\begin{figure}[htp]
\centering	
	\includegraphics[width=3in,keepaspectratio]{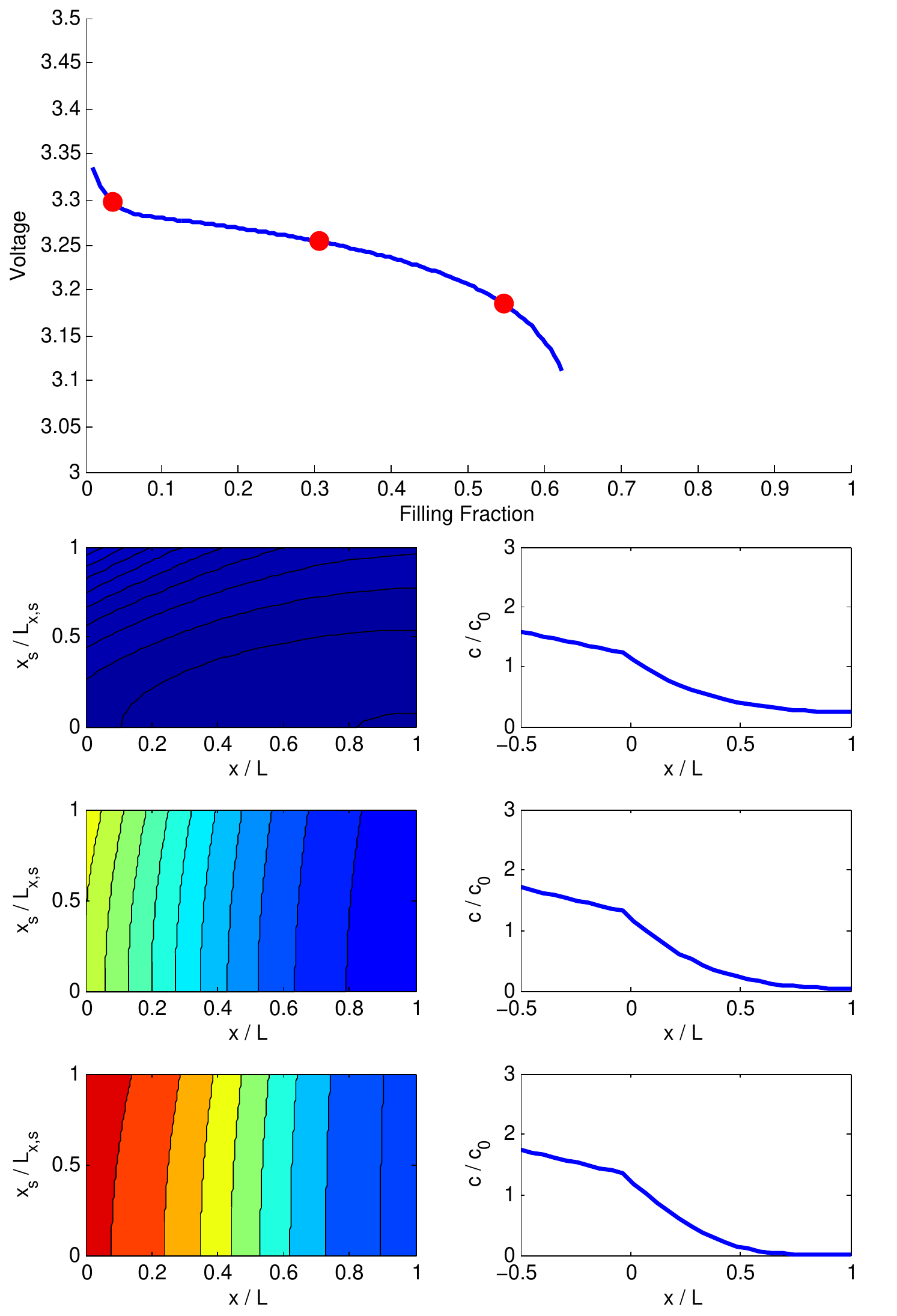}
\caption{\textbf{Depletion of the electrolyte at higher current}.  This figure shows the depletion of the electrolyte accompanying Figure \ref{homogvarcurr1} for the 15C discharge.  The left figure shows the solid concentration while the right figure demonstrates the electrolyte concentration profile in the separator and electrode.}
\label{homogelectdep1}
\end{figure}

As the current is increased, gradients in solid concentration across the electrode begin to become prevalent.  At the same time, transport limitations in the electrolyte lead to a capacity limitation, as the electrolyte is incapable of delivering lithium quickly enough deeper into the electrode.  Figure \ref{homogelectdep1} demonstrates the electrolyte depletion leading to the concentration polarization in the 15C discharge curve.   While the voltages appear to stop, these are actually points where it drops off sharply.  Tighter tolerances, which can significantly increase the computation time, are needed to get the voltage down to zero.

It is important to note that $\delta_d$ is not the ratio of diffusivities, but the ratio of diffusive times.  Therefore, as particle size increases, the diffusive time scales as the square of the particle size.  Solid diffusivities are typically much slower than in the electrolyte.  To demonstrate the effect of increased current with slower solid diffusion, Figure \ref{homogvarcurrhighdiff1} demonstrates the same discharge rates as the previous figure, except the solid diffusive time scale has been increased to 100 times the electrolyte diffusive time scale.  

For decreased solid diffusivity, concentration gradients in the depth direction of the particles are more prevalent.  At low current (i.e. slow discharge), the gradients in the electrode and particles are minimal.  As the current is increased, gradients in the particles begin to emerge.  At the fastest discharge rate, these solid concentration gradients become very large.  Finite volume effects at the surface of the particles increase the overpotential substantially, producing a sharp voltage drop-off and low utilization.  This effect is caused by the slow solid diffusion only. Despite plenty of lithium being available in the electrolyte, high surface concentrations block available sites for intercalation.

To show the effect of solid diffusion alone, Figure \ref{homogvardiff1} demonstrates the effect of decreasing solid diffusivity at a constant discharge rate.  When the diffusive time scales of the solid and electrolyte are comparable, each particle fills homogeneously.  There are small variations along the length of the electrode, but these do not affect the utilization, as almost 100\% of the electrode is utilized.  

As the solid diffusivity is decreased, and the diffusive time scale approaches 50 times the electrolyte diffusive time scale, we see over a 10\% drop is capacity.  Concentration gradients in the solid particles begin to emerge.  As the solid diffusivity is further decreased, and the solid diffusive time scale approaches 100 times the electrolyte diffusive time scale, the solid concentration gradients become quite large, leading to a 50\% drop in capacity.  While these changes in $\delta_d$ seem significant, they represent approximately a two order of magnitude change in diffusivity, and a one order of magnitude change in particle size.

\subsection{Phase Separating Particles}

For the case of phase separating materials, the equilibrium homogeneous voltage curve is non-monotonic.  This is demonstrated in Figure \ref{ocpcomp1}, for regular solution parameters greater than 2$k_BT$.  For these materials, the free energy curve has two local minima.  When the second derivative of the free energy with respect to filling fraction changes sign (positive to negative), the system is unstable for infinitesimal perturbations, resulting in phase separation.  A tie line represents the free energy of the system, and the proportion of the two phases changes as the system fills.  

Modeling phase separating materials requires the use of the Cahn-Hilliard free energy functional as given in Equation (\ref{cahnhilliardFE1}), and the Cahn-Hilliard diffusional chemical potential, given in Equation (\ref{cahnhilliardCPot1}).  When we insert the chemical potential into the modified Butler-Volmer Equation, we obtain a forced Allen-Cahn type equation.  Here, we present the first solution of multiple phase separating particles in a porous electrode.

\begin{figure}[htp]
\centering	
	\includegraphics[width=3in,keepaspectratio]{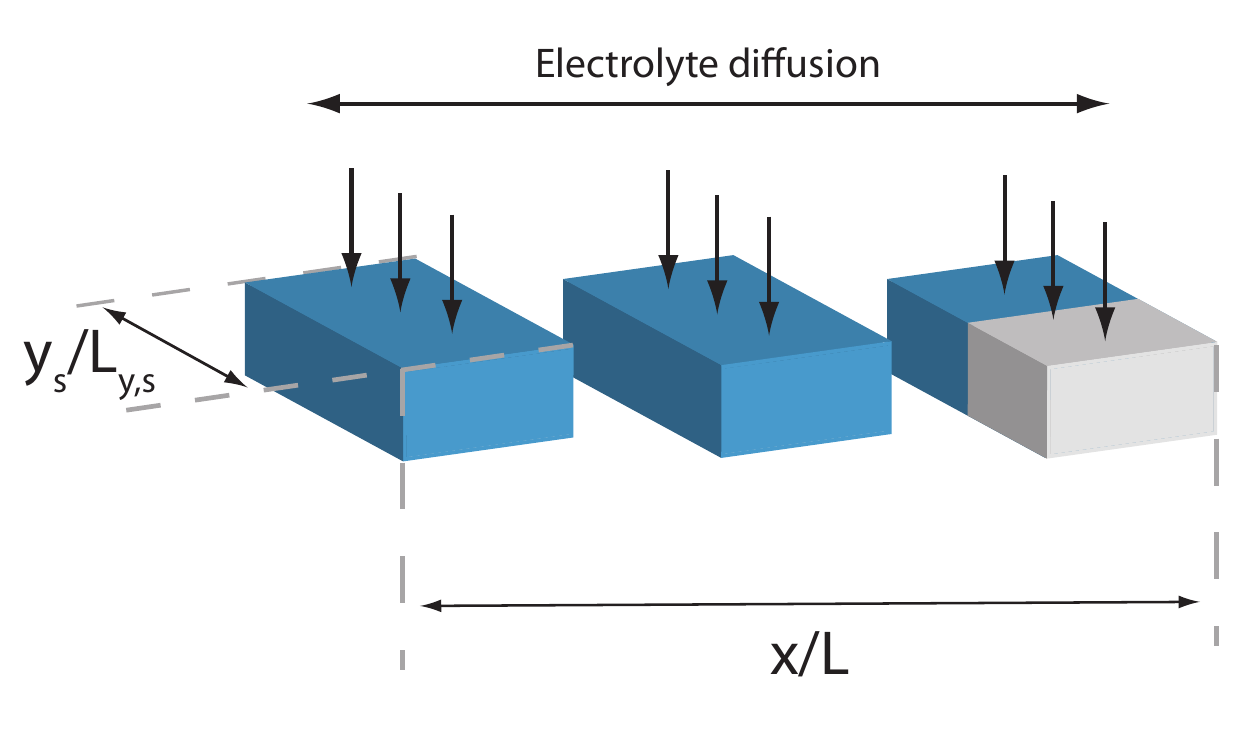}
\caption{\textbf{Plot axes for reaction-limited phase separating particles}.  This figure shows how the results are plotted below for porous electrodes with reaction-limited phase separating nanoparticles.  The y-axis of the contour plots represent the length along the surface of the particle, since diffusion is assumed to be fast in the depth direction.  The x-axis represents the depth in the electrode.}
\label{psdiag1}
\end{figure}

\begin{figure}[htp]
\centering	
	\includegraphics[width=3in,keepaspectratio]{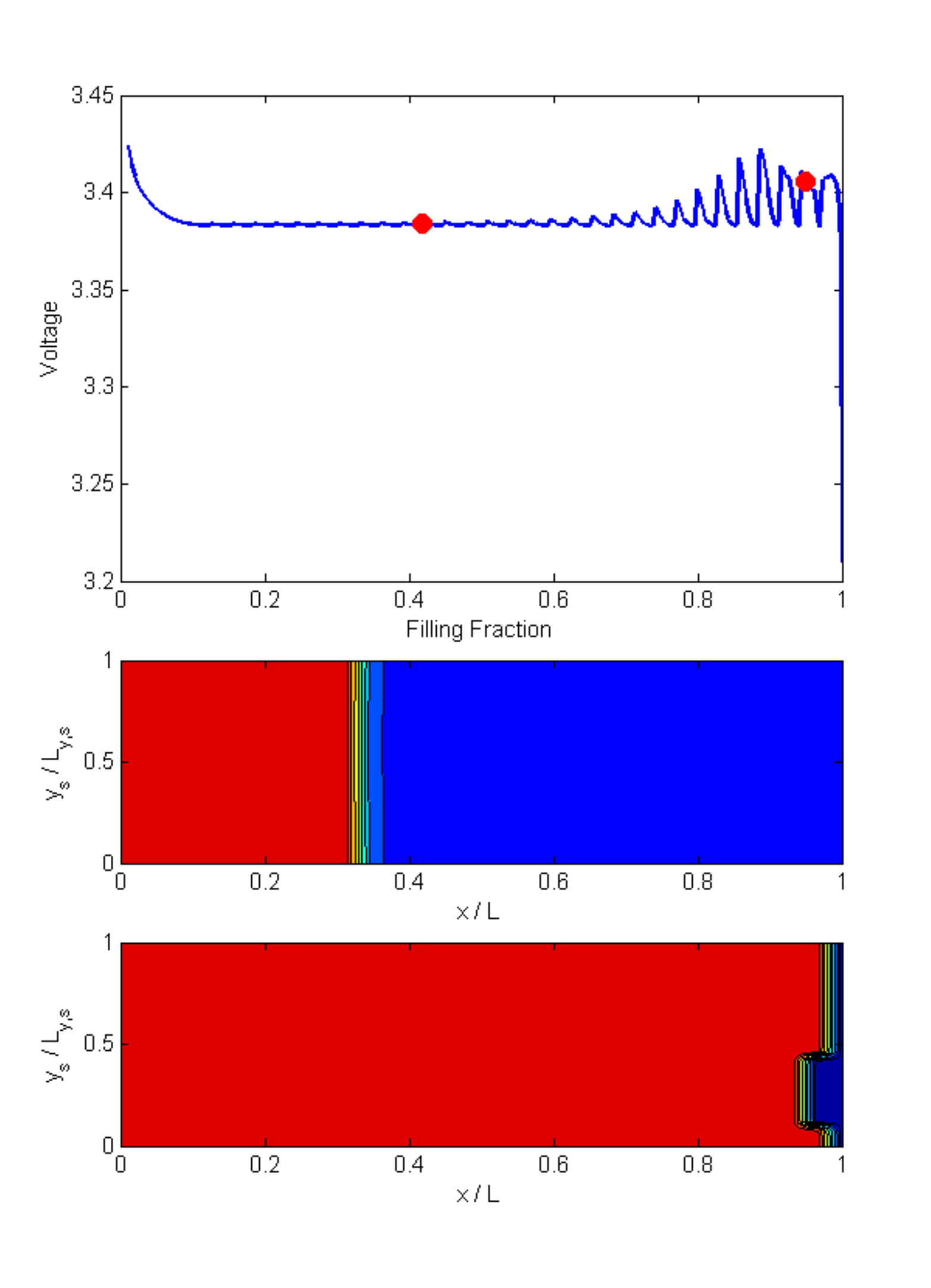}
\caption{\textbf{Phase separating particles slowly discharged}. This figure shows slow discharge (approx. C/30) of phase separating particles.  Adequate electrolyte diffusion and discrete filling don't allow time for the particles to phase separate early on.  At the end of the discharge, sufficient time allows the particles to phase separate.}
\label{chr_ps_1}
\end{figure}

For phase separating particles, values of $\Omega=4k_BT$ and $\tilde{\kappa}=0.001$ were used along with a regular solution model to model the homogeneous chemical potential, $\overline{\mu}$.  The same exchange current as above was used.  The figures are similar to those of the homogeneous plots, but instead of the depth direction, we now plot along the surface.  Figure \ref{psdiag1} depicts the axes plotted.  This assumes that the diffusion into the particle is fast, and that the process is essentially surface reaction limited.  This is a reasonable approximation for LiFePO$_4$. \cite{bai2011}  Figure \ref{chr_ps_1} demonstrates slow discharge (approx. C/30).  

Initially, the discrete filling of the electrode suppresses phase separation inside the particles.  Towards of the end of the discharge, decreased electrolyte diffusion (from longer path length) allow for particles to phase separate.  Another important feature of the simulation is the voltage spikes towards the end of the simulation.  These voltage spikes, which are on the order of the thermal voltage, are an artifact of the discrete nature of the model.  Towards the end of the simulation, only a few particles remain to fill, therefore the voltage is dominated by effectively the single particle response.  Dreyer \emph{et al.} demonstrated this previously for phase separating particles filling homogeneously. \cite{dreyer2010,dreyer2011}  

\begin{figure}[htp]
\centering	
	\includegraphics[width=3in,keepaspectratio]{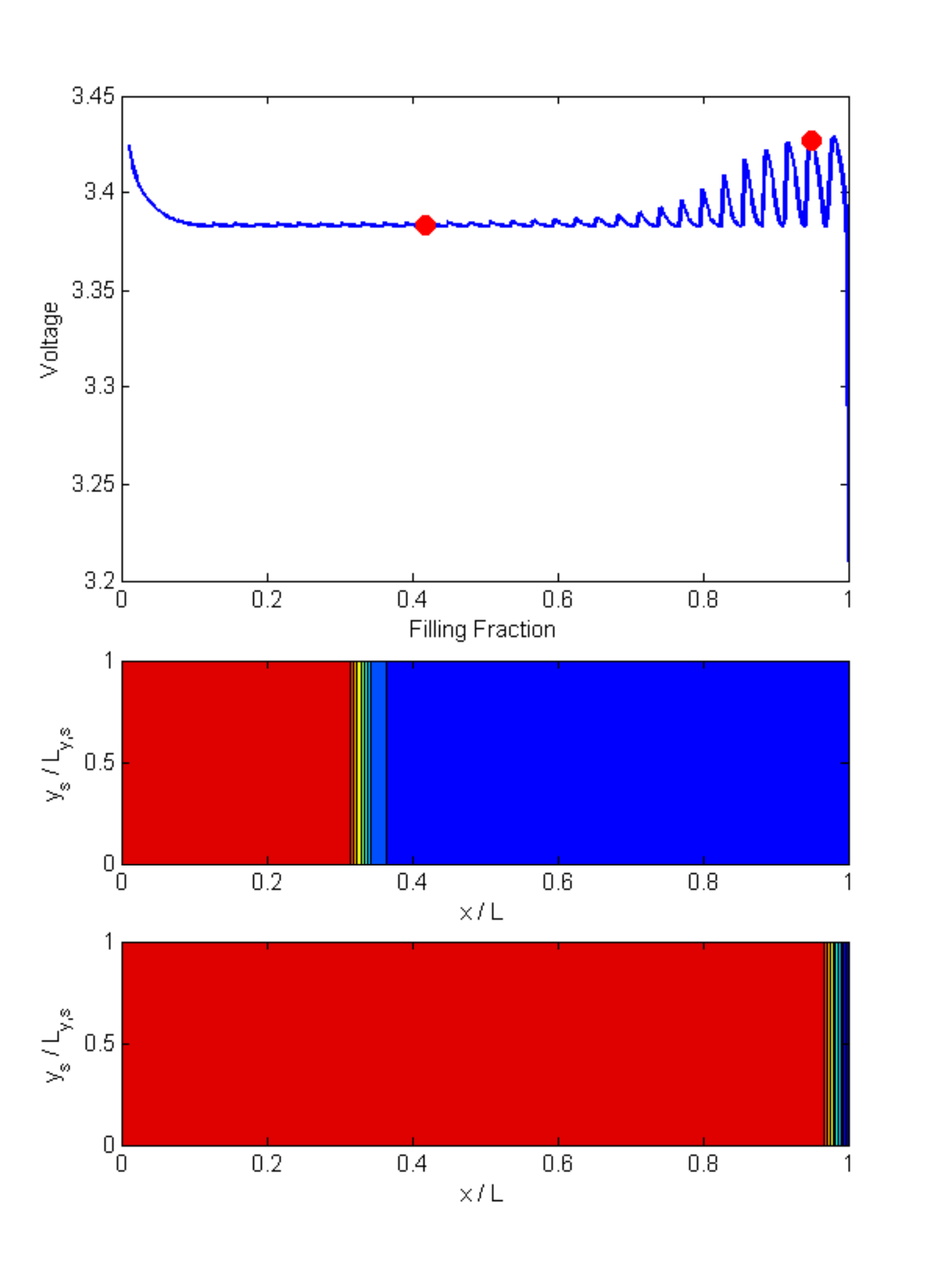}
\caption{\textbf{Phase separating particles including coherent stress effects slowly discharged}. This figure shows slowly discharge (approx. C/30) phase separating particles.  The inclusion of the coherent stress effects suppresses phase separation inside the particles.  This figure is the same as Figure \ref{chr_ps_1}, with an additional coherent stress term.}
\label{chr_ps_se_1}
\end{figure}

The kinetics of phase separating particles can also be heavily influenced by stress effects, as demonstrated recently by Cogswell and Bazant. \cite{cogswell2012} Including stress involves the addition of energy terms in the free energy model.
With stress included, the full form of the bulk free energy functional is
\begin{equation}
G[\tilde{c}(x)] = \int_V\left[\rho_s f(\tilde{c})+\frac{1}{2}\kappa\left(\nabla \tilde{c}\right)^2 +\frac{1}{2}C_{ijkl}\varepsilon_{ij}\varepsilon_{kl} - \overline{\sigma}_{ij} \overline{\varepsilon}_{ij}\right]dV,
\label{freeenergystress1}
\end{equation}
where the additional terms represent the elastic strain energy and the homogeneous component of the total strain, respectively. (Here, we neglect the surface term, which mainly affects nucleation of phase separation via surface wetting~\cite{bai2011}.)   The effects of  coherency strain on phase separation can be approximated by a volume averaged stress term \cite{cahn1961,cahn1962cohnuc,cahn1962sd}  The homogeneous component of the total strain is then 
\begin{equation}
\frac{1}{2}C_{ijkl}\varepsilon_{ij}\varepsilon_{kl}\approx \frac{1}{2}B\left(\tilde{c}-X\right)^2,
\label{homstrainapp1}
\end{equation}
where $X$ is the volume averaged concentration.  This approximation limits local fluctuations and promotes homogeneous filling depending on the value of the constant $B$ (which generally depends on orientation~\cite{cogswell2012}).  Including this term the chemical potential we obtain
\begin{equation}
\mu = \overline{\mu}-\nabla\cdot\left(\frac{\kappa}{\rho_s}\nabla \tilde{c}\right) + \frac{B}{\rho_s}\left(\tilde{c}-X\right).
\label{ch_with_se1}
\end{equation}
As the difference between the local and average global particle concentration increases, the overpotential required to drive the intercalation reaction increases.  This promotes homogeneous filling of the particles.  Figure \ref{chr_ps_se_1} demonstrates how this additional term suppresses phase separation.  However, the discrete filling still produces the voltage plateau and spikes in voltage.

While these spikes appear to be large, they are actually on the order of the thermal voltage or smaller.  At typical voltage scales (2.0V-3.5V) these spikes are not seen, resulting in a flat voltage profile as seen in experimental data for LiFePO$_4$.  This demonstrates that a phase separating material's flat voltage profile can be modeled without modeling phase transformation itself.  The voltage spikes depend on the value of the Damk\"{o}hler number, or ratio of the diffusion time across the porous electrode to the typical reaction time to fill an active particle.  

\begin{figure}[htp]
\centering	
	\includegraphics[width=3in,keepaspectratio]{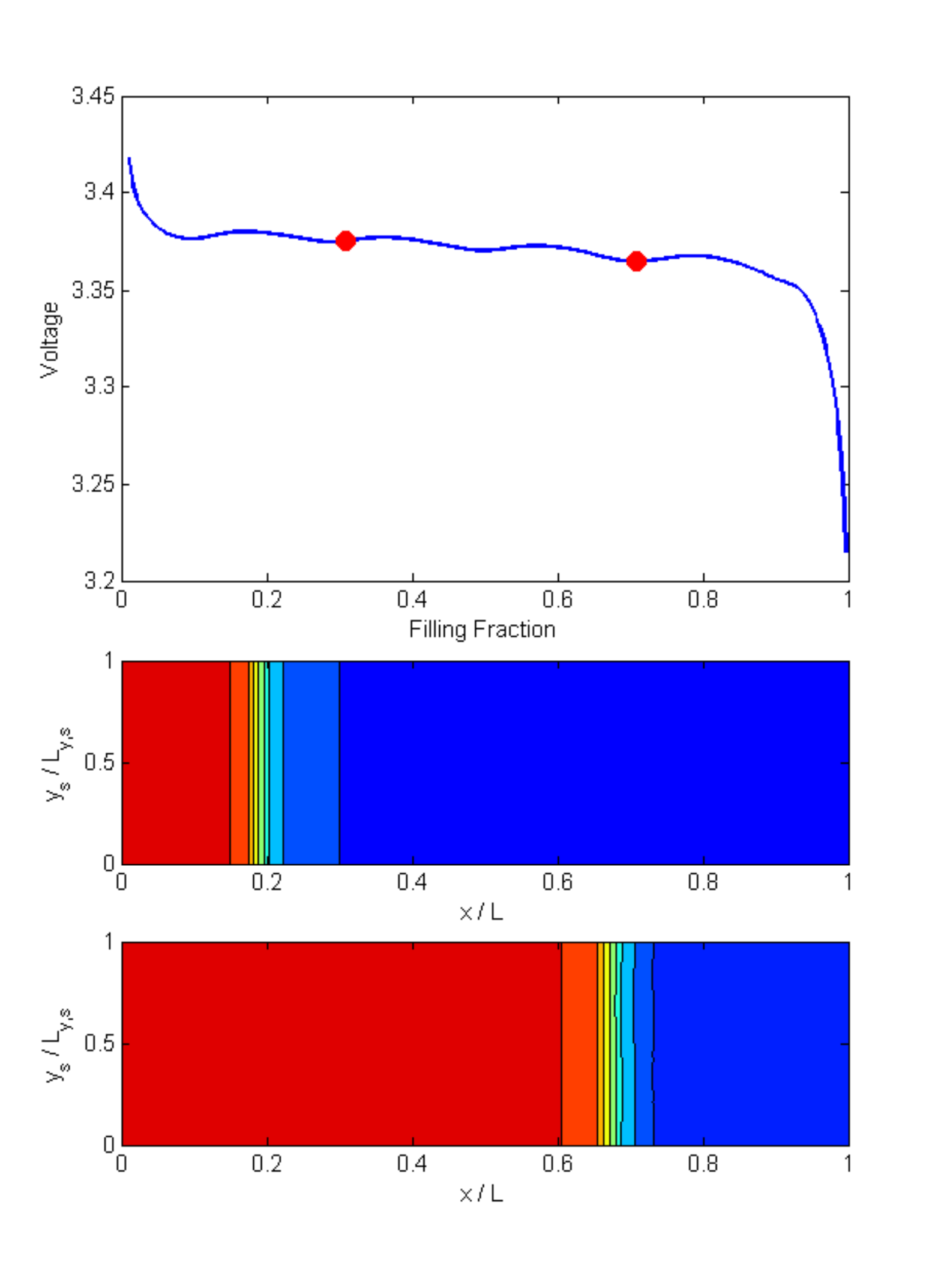}
\caption{\textbf{Effect of current on phase separating particles}.  When discharged at a higher C-rate (in this example, 3C), the size of the discrete particle filling is larger, leading to more particles filling simultaneously and a voltage curve that resembles solid solution behavior.}
\label{chrhighcurr1}
\end{figure}

\begin{figure}[htp]
\centering	
	\includegraphics[width=3in,keepaspectratio]{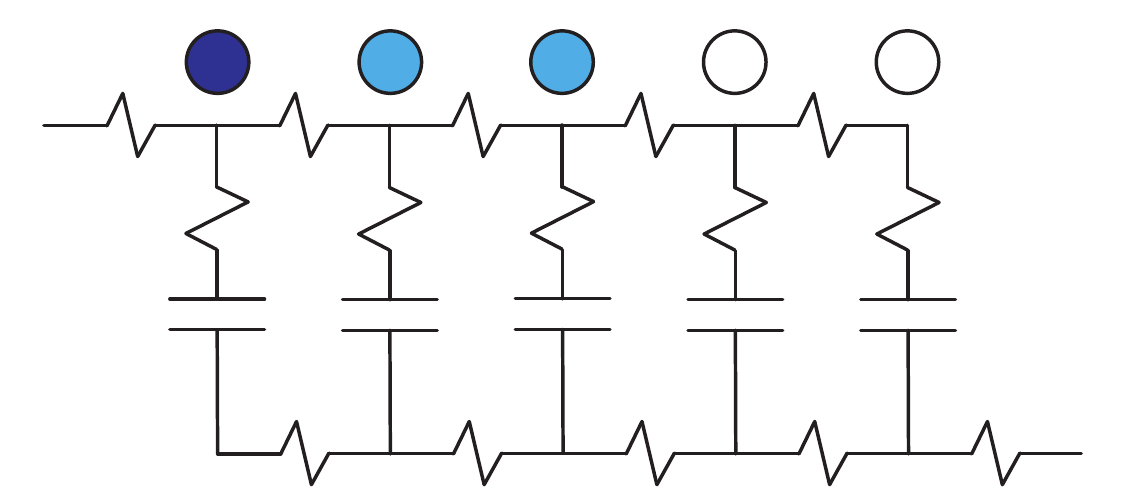}
\caption{\textbf{Equivalent circuit model for a porous electrode}.  This equivalent circuit represents a typical porous electrode in cases without significant electrolyte depletion, where the pore phase maintains nearly uniform conductivity.  Resistors represent the contact, transport, and charge transfer resistances, and the capacitance of the particles is represented by a capacitor.  All elements are not necessarily linear.}
\label{peequivcirc1}
\end{figure}

Figure \ref{chrhighcurr1} shows a faster (3C) discharge of the phase separating particles.  The voltage spikes are suppressed and the voltage curve resembles ``solid solution" behavior.  There are three small voltages fluctuations present in the simulation which are caused by the discrete filling effect.  However, instead of individual particles filling, now larger clusters of particles fill to alleviate the current (i.e. the number of active particles, or particles undergoing intercalation, has increased).  To explain this, consider the equivalent circuit for a porous electrode in Figure \ref{peequivcirc1}. 

The particles are represented by equivalent circuits.  Each particle (which could also be considered to be a cluster of particles with similar properties) has a charge transfer resistance, $R_{ct}$, and capacitance $C_p$.  These values can be non-linear, and vary depending on the particle filling fraction and/or local potential.  For each particle or cluster of particles, there is a charging time, $\tau_c$, which scales as
\begin{equation}
\tau_c \sim R_{ct}C_p.
\label{chargetime1}
\end{equation}
For a given discharge rate at constant current, particles in the electrode must alleviate a given amount of lithium per time in the electrode.  

The number of active particles scales as 
\begin{equation}
n_{ap} \sim \tau_c\tilde{I}.
\label{numactivepart1}
\end{equation}
As the discharge rate is increased, the number of active particles increases until it spans the electrode, resulting in the electrode filling homogeneously.  For fast kinetics or slow discharge rates, the number of active particles is small, which produces the discrete filling effect.  For the non-monotonic OCP of homogeneous phase separating particles, the voltage plateau has three filling fractions that can exist in equilibrium: the left miscibility gap filling fraction, half filling fraction, and right miscibility gap filling fraction.  As the particles fill, if the kinetics are sufficiently fast, then other particles close to the active particle will empty to reach the equilibrium voltage (the plateau voltage).  This increase in voltage for each particle as it deviates from the voltage at the spinodal concentration leads to an increase in cell voltage, producing the voltage spikes.

For slower kinetics, this effect is suppressed by two mechanisms.  First, the charge transfer resistance is larger, leading to higher charging times and subsequently a larger number of active particles.  Also, slower kinetics hinders the ability of particles to easily insert/remove lithium, which prevents the particles from emptying and increasing the voltage, leading to the spikes.  

\section{Summary}

In this paper, we have generalized porous electrode theory using principles of non-equilibrium thermodynamics.  A unique feature is the use of the variational formulation of reaction kinetics~\cite{acr2012,10.626}, which allows the use of phase field models to describe macroscopic phase transformations in porous electrodes for the first time.  
The thermodynamic consistency of all aspects of the model is crucial. Unlike existing battery simulation models, the open circuit voltage, reaction rate, and solid transport properties are not left free to be independently fit to experimental data. Instead, these properties are all linked consistently to the electrochemical potentials of ions and electrons in the different components of the porous electrode. Moreover, emergent properties of a phase-separating porous electrode, such as its voltage plateau at low current, are not fitted to empirical functional forms, but rather follow from the microscopic physics of the material. This allows the model to capture stochastic, discrete phase transformation events, which are beyond the reach of traditional diffusion-based porous electrode theory. 

In a companion paper~\cite{ferguson2012b}, we will apply the model to predict the electrochemical behavior of composite, porous graphite anodes~\cite{harris2010} and LFP cathodes~\cite{dreyer2010}, each of which have multiple stable phases. Complex nonlinear phenomena, such as narrow reaction fronts, mosaic instabilities, zero current voltage gap, and voltage fluctuations, naturally follow from the simple physics contained in the model. The model is able to fit experimental data for phase transformations in porous electrodes under very different conditions, limited either by electrolyte diffusion~\cite{harris2010} or by reaction kinetics~\cite{dreyer2010}.

\vskip 0.1in
This work was supported by the National Science Foundation under Contracts DMS-0842504 and DMS-0948071 (H. Warchall) and by a seed grant from the MIT Energy Initiative.

\section{List of symbols used}

\emph{NOTE: unless explicitly noted, all quantities with a tilde denote dimensionless quantities.  Energies are scaled by the thermal energy, $k_BT$, and potentials are scaled by the thermal voltage, $k_BT/e$}

\textbf{Symbols used:}

\begin{supertabular}{ l l }
$a$ & activity (dimensionless) \\
$a_p$ & pore area per volume [1/m] \\
$A$ & area [m$^2$] \\
$A_{cell}$ & area of unit cell (CST derivation) [m$^2$] \\
$A_{sep}$ & area of separator [m$^2$] \\
$B$ & volume averaged elastic strain energy [J/m$^3$] \\
$c$ & number concentration [1/m$^3$] \\
$\tilde{c}$ & dimensionless concentration \\
$\overline{c}$ & volume averaged number concentration [1/m$^3$] \\
$c_{max}$ & maximum number concentration (solubility limit) [1/m$^3$] \\
$C_p$ & capacitance [C/V] \\
$C_{ijkl}$ & elastic stiffness tensor [J/m$^3$] \\
$d$ & dimensionality \\
$D$ & diffusivity [m$^2$/s] \\
$D_{amb}$ & ambipolar diffusivity [m$^2$/s] \\
$D_{chem}$ & chemical diffusivity [m$^2$/s] \\
$D_o$ & tracer diffusivity [m$^2$/s] \\
$D_p$ & diffusivity inside a pore [m$^2$/s] \\
$\overline{D}$ & effective diffusivity [m$^2$/s] \\
$e$ & elementary charge [C] \\
$\textbf{e}_i$ & coordinate vector \\
$E_O$ & reference energy of oxidant [J] \\
$E_R$ & reference energy of reductant [J] \\
$E_\ddagger$ & reference energy of transition state [J] \\
$f$ & homogeneous free energy per volume [J/m$^3$] \\
$\textbf{F}$ & number flux [1/m$^2$s] \\
$g$ & free energy per lattice site [J] \\
$G$ & total free energy [J] \\
$\textbf{i}$ & current density [C/m$^2$s] \\
$i_o$ & exchange current density [C/m$^2$s] \\
$I$ & total current [C/s] \\
$j_{in}$ & reaction flux [1/m$^2$s] \\ 
$k_o$ & rate constant [1/s] \\
$k^o$ & modified rate constant [1/s] \\
$k_B$ & Boltzmann's constant [J/K] \\
$L$ & characteristic length [m] \\
$L_p$ & characteristic pore length \\
$M$ & mobility [m$^2$/Js] \\
$M_i$ & chemical symbol of species $i$ \\
$n$ & number electrons transferred \\
$n_{ap}$ & number of active particles \\
$\textbf{N}$ & number species flux [1/m$^2$s] \\
$P_L$ & loading percent of active material by volume \\
$q$ & species charge number \\
$r$ & radial direction [m] \\
$R$ & reaction rate [1/m$^3$s] \\
$R_{ct}$ & charge transfer resistance \\
$S_1$ & stoichiometric sum of reactants \\
$S_2$ & stoichiometric sum of products \\
$s_i$ & stoichiometric coefficients \\
$t$ & time [s] \\
$t_d$ & characteristic diffusion time [s] \\
$t_p$ & percolation exponent \\
$t_\pm$ & transference number of positive/negative species \\
$T$ & temperature [K] \\
$V$ & volume [m$^3$] \\
$x$ & spatial direction [m] \\
$X$ & average dimensionless concentration \\
$z_i$ & charge number of species $i$ \\
\end{supertabular}

\textbf{Greek symbols:}

\begin{supertabular}{ l l }
$\alpha$ & transfer coefficient \\
$\beta$ & ratio of solid:electrolyte lithium capacity \\
$\delta_d$ & ratio of characteristic solid:electrolyte diffusive times \\
$\epsilon$ & porosity (pore volume per total volume) \\
$\varepsilon_{ij}$ & strain \\
$\overline{\varepsilon}_{ij}$ & homogeneous component of elastic strain \\
$\eta$ & overpotential [V] \\
$\eta$ & dimensionless overpotential \\
$\gamma$ & activity coefficient [m$^3$] \\
$\gamma_\ddagger$ & activity coefficient of transition state [m$^3$] \\
$\kappa$ & gradient energy [J/m] \\
$\tilde{\kappa}$ & dimensionless gradient energy \\
$\mu$ & chemical potential [J] \\
$\tilde{\mu}$ & dimensionless chemical potential \\
$\mu^{ex}$ & excess chemical potential [J] \\
$\overline{\mu}$ & homogeneous chemical potential [J] \\
$\mu^o$ & reference chemical potential [J] \\
$\nu$ & attempt frequency [1/s] \\
$\Omega$ & regular solution interaction parameter [J] \\
$\phi$ & electrolyte potential [V] \\
$\tilde{\phi}$ & dimensionless potential \\
$\Phi$ & volume fraction \\
$\rho_s$ & site density [1/m$^3$] \\
$\sigma$ & conductivity [S/m] \\
$\overline{\sigma}$ & effective conductivity [S/m] \\
$\overline{\sigma}_d$ & diffusive mean conductivity [m$^2$/s] \\
$\overline{\sigma}_{ij}$ & applied external stress tensor [N/m$^2$] \\
$\tau$ & time between transitions [s] \\
$\tau_c$ & charging time [s] \\
$\tau_p$ & tortuosity (pore length per total length) \\
$\tau_o$ & barrier-less transition time [s] \\
\end{supertabular}

\textbf{Subscripts:}

\begin{supertabular}{ l l }
$_+$ & positive species \\
$_-$ & negative species \\
$_B$ & Bruggeman model \\
$_c$ & critical (percolation model) \\
$_{eq}$ & equilibrium \\
$_i$ & species $i$ \\
$_O$ & oxidant \\
$_p$ & pore phase \\
$_{perc}$ & percolation model \\
$_M$ & metal/electron conducting phase \\
$_{max}$ & maximum \\
$_{min}$ & minimum \\
$_R$ & reductant \\
$_s$ & solid (intercalation) phase \\
\end{supertabular}

\textbf{Superscripts:}

\begin{supertabular}{ l l }
$^B$ & Bruggeman model \\
$^{HS}$ & Hashin-Shtrikman model \\
$^{perc}$ & Percolation model \\
$^{Wiener}$ & Wiener model \\
\end{supertabular}

\clearpage

\bibliography{elec35}

\end{document}